\documentclass[twocolumn]{aa} 
\usepackage{graphicx}
\usepackage{txfonts}
\usepackage{caption}
\usepackage{subcaption}

\begin{document}

\title{Photometry and model of near-Earth asteroid 2021 DW1\\ from one apparition}
\titlerunning{Photometry and model of 2021 DW1}
\authorrunning{Kwiatkowski et al.}

\author{T.~Kwiatkowski\inst{1} \and P.~Kole\'nczuk\inst{1} \and A.~Kryszczy\'nska\inst{1} \and D.~Oszkiewicz \inst{1} \and K.~Kami\'nski\inst{1} \and M.~K. Kami\'nska\inst{1} \and V.~Troianskyi\inst{1,3} \and B.~Skiff\inst{2} \and N.~Moskowitz\inst{2}  \and V.~Kashuba\inst{3} \and  M.-J.~Kim\inst{4} \and T.~Kim\inst{5} \and S.~Mottola\inst{6} \and T.~Santana-Ros\inst{7,8} T.~Kluwak\inst{9} \and L.~Buzzi\inst{10} \and P.~Bacci\inst{11} \and P.~Birtwhistle\inst{12} \and R.~Miles\inst{13} \and J.~Chatelain\inst{14}
}

\institute{Astronomical Observatory Institute, Faculty of Physics, A. Mickiewicz University, S{\l}oneczna 36, 60-286 Pozna{\'n}, Poland \label{inst1} 
\and Lowell Observatory, 14000 W Mars Hill Rd, 86001 Flagstaff, AZ, USA \label{inst2}
\and Astronomical Observatory of Odessa I.I.Mechnikov National University, Odessa, Ukraine\label{inst3}
\and Korea Astronomy \& Space Science Institute, 776 Daedeok-daero, Yuseong-gu, Daejeon 34055, Republic of Korea\label{inst4}
\and National  Youth  Space  Center,  Goheung,  Jeollanam-do,  59567, Korea\label{inst5}
\and Deutsches Zentrum f\"or Luft- und Raumfahrt (DLR), Institute of Planetary Research, Berlin, Germany\label{inst6}
\and Departamento de F\'isica, Ingenier\'ia de Sistemas y Teor\'ia de la Se\~nal, Universidad de Alicante, Alicante, Spain\label{inst7}
\and Institut de Ciencies del Cosmos (ICCUB), Universitat de Barcelona, Barcelona, Spain\label{inst8}
\and Platanus Observatory (IAU code K80), Lus{\'o}wko, Poland \label{inst9}
\and "G.V.Schiaparelli" Astronomical Observatory, Varese, Italy\label{inst10}
\and Osservatorio di San Marcello Pistoiese, GAMP Gruppo Astrofili Montagna Pistoiese, Italy\label{11}
\and Great Shefford Observatory, Berkshire, United Kingdom\label{12}
\and British Astronomical Association, Burlington House, Piccadilly, London, United Kingdom\label{13}
\and Las Cumbres Observatory, Goleta, CA  93117, USA\label{14}
}
         
\date{\today} 

\offprints{T. Kwiatkowski, e-mail: tkastr at vesta.astro.amu.edu.pl}

\date{Received xx xx xxxx / Accepted xx xx xxxx}

 
\abstract
{}
{Very Small Asteroids (objects with diameters smaller than about 150~m) can
be spun-up by the YORP effect to rotation periods as short as tens of
seconds. This effect has been observed for many of them. It is also 
hypothesized, that in the same process their spin axes are asymptotically drawn 
to the position perpendicular to the orbital plane. So far this effect has
been observed only for one VSA and needs further verification. For that,
spin axes of several other VSAs should be determined by observing their
brightness variations at many different positions on the sky.}
{On 4~March 2021 at 9~UTC a 30-m in diameter near-Earth asteroid
2021~DW$_1$ passed the Earth at a distance of 570~000~km, reaching the
maximum brightness of $V=14.6\,\mathrm{mag}$.  We observed it
photometrically from 2~March, when it was visible at $V=16.5\,\mathrm{mag}$,
until 7~March ($V=18.2\,\mathrm{mag}$).  During that time 2021~DW$_{1}$
swept a $170\degr$ long arc in the northern sky, spanning solar phase angles
in the range from $36\degr$ to $86\degr$. This made it an excellent target
for physical characterisation, including spin axis and shape derivation.}
{Convex inversion of the asteroid lightcurves gives a sidereal period of
rotation  $P_{\mathrm{sid}}=0.013760 \pm 0.000001\,\mathrm{h}$, and two 
solutions for the spin axis
ecliptic coordinates: (A) $\lambda_1=57\degr \pm 10\degr, \beta_1=29\degr
\pm 10\degr$, and (B) $\lambda_2=67\degr \pm 10\degr, \beta_2=-40\degr \pm
10\degr$.  
The magnitude-phase curve can be
fitted with a standard H,~G function with $H=24.8\pm 0.5\,\mathrm{mag}$ and
an assumed $G=0.24$.  The asteroid colour indices are $g-i=0.79\pm
0.01\,\mathrm{mag}$, and $i-z=0.01\pm 0.02\,\mathrm{mag}$ which indicates an
S taxonomic class, with an average geometric albedo $p_V=0.23\pm 0.02$.  The
asteroid effective diameter, derived from $H$ and $p_V$, is
$D_\mathrm{eff}=30\pm 10\,\mathrm{m}$.}
{It was found that the inclination of the spin axis of 2021~DW$_1$ is not 
perpendicular to the orbital plane (obliquity $\epsilon=54\degr\pm 10\degr$ or
$\epsilon=123\degr\pm 10\degr$). More spin axes of VSAs should be determined
to check, if 2021~DW$_1$ is an exception or a typical case.}


\maketitle
%

\section{Introduction}

Very Small Asteroids (VSAs) are objects with diameters $D<150$~m.  They
often rotate with periods shorter than 2~h enabling us to study their
internal structure by comparing the centrifugal force with the material
forces holding them together \citep{Holsapple2007}. Because of their small
sizes, VSAs are sensitive to
the YORP effect \citep{Rubincam2000}, which is a torque
induced on the rotating body by the thermal radiation emitted by its
surface complemented by a torque produced
by scattered sunlight.  YORP can either spin up or slow down the asteroid rotation as well
as change the obliquity of its spin axis $\epsilon$, which is an angle
between the normal to the asteroid orbital plane and its rotation axis. 
While the fast rotation has been observed for many VSAs, their spin axes
were not determined except for one object\footnote{To be exact, in the 
DAMIT database \citep{Durech+2010} there are two other VSAs (2008~TC3
and 2012~TC4), for which spin
axes have been determined, but both of them are non-principal axis rotators
and show the effect of tumbling}: (54509) YORP (which name is the
same as the name of the effect itself). (54509) was the first
asteroid for which the effect of YORP has been observed \citep{Lowry+2007,
Taylor+2007}. The obliquity of the (54509) spin
axis is $\epsilon=173\degr$ which means it is nearly perpendicular to the
asteroid orbital plane.  Such orientation of the spin axis was found as an end state
of the YORP evolution in the simulations performed by \citet{Capek+2004}
for objects with finite surface thermal conductivity. If
their prediction is true, then for the VSAs, which experienced a strong YORP
effect for a long time (and the fastest rotating VSAs are such
objects) we should observe spin axis obliquities close to $0\degr$ or
$180\degr$. Recently \citet{Golubov+2021} have shown, that for very small 
objects of highly irregular shape, the transverse heat conduction (TYORP) can
add new asymptotic states for the obliquity. For this reason it would be
interesting to verify those predictions with observations of VSAs.
To do that, we should observe their
lightcurves at many different positions on the sky to be able to determine 
their spin axes. For NEAs this condition is met either by the 
Earth co-orbital asteroids -- and (54509) is an example of such 
objects -- or by objects which
during their close encounter with the Earth can be observed along a long arc
on the sky.

The main goal of this paper is to present the observations and modelling
of the near-Earth asteroid 2021~DW$_1$, which allowed us to derive its spin
axis. Apart from that, we were able to determine many other physical
properties of this VSA, such as its rotation period, shape, size, and
taxonomy class, adding new data to this still poorly characterised group of
objects. 

\begin{figure}[!t]
\resizebox{\hsize}{!}{\includegraphics[clip]{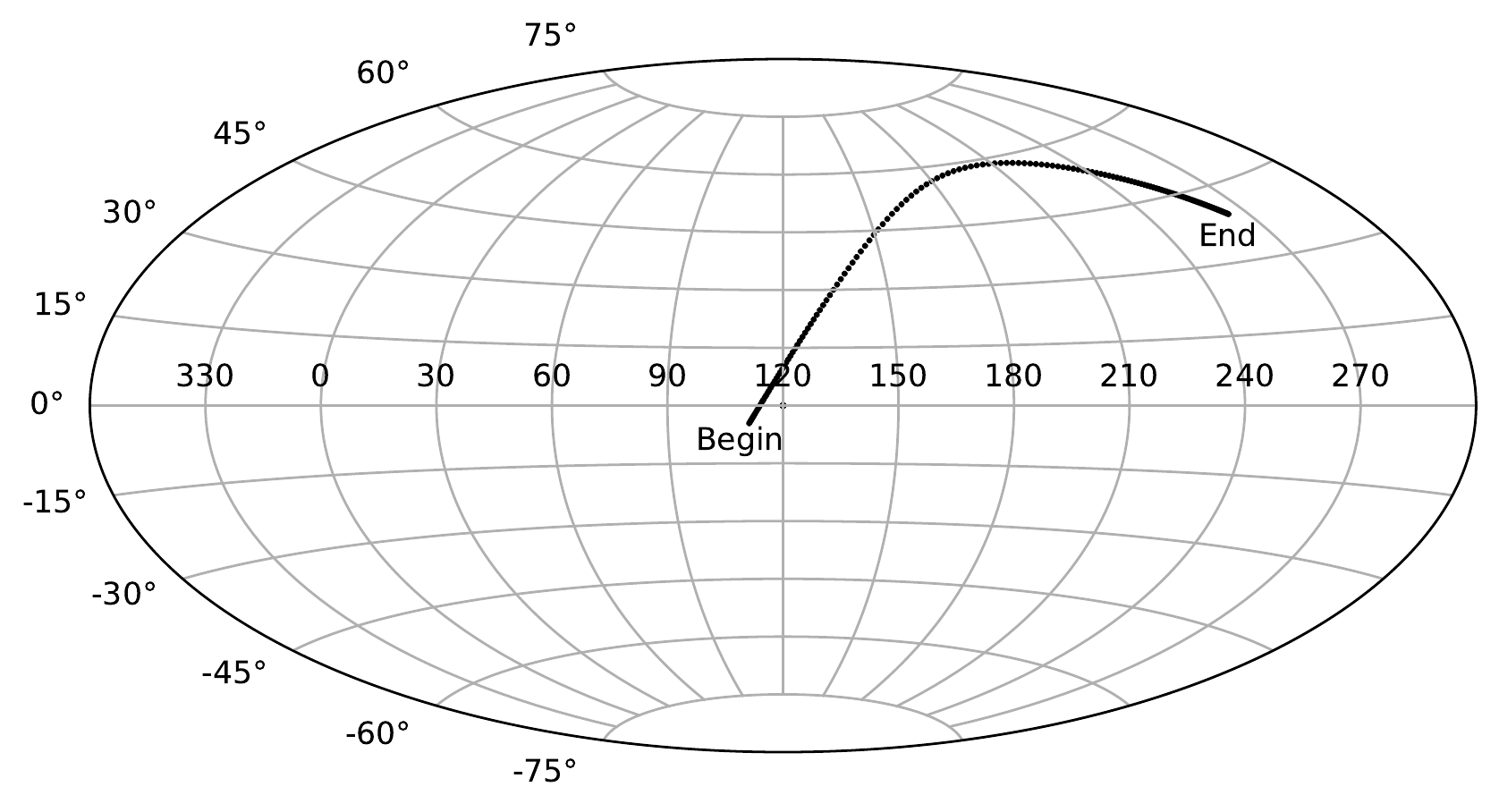}}
\caption{A trail made by 2021~DW$_1$ on the sky during our observing
campaign, plotted in the RA, Dec equatorial coordinates.  The beginning of
the arc refers to 2~Mar, 4~UTC, and the end to 7~Mar, 7~UTC.  The length of
the trail is about $170\degr$.  Note that two observations made on 7~Mar
were finally not used for lightcurve inversion due to high photometric noise.}
\label{PathInSky}
\end{figure}

\section{Observations and data reduction}

On 16~February 2021 a near-Earth asteroid was discovered by the Pan-STARRS~1 
survey at Haleakala, Hawaii.  The discovery was reported in MPEC~2021-D73
circular\footnote{https://www.minorplanetcenter.net/mpec/K21/K21D73.html}
and the object was designated 2021~DW$_{1}$.  Its first estimate of the 
absolute magnitude gave a value of $H=25$~mag, which translated roughly 
to an effective diameter of 40~m, indicating it is a VSA with 
a rotation period possibly  shorter than $2.2\,\mathrm{h}$.  
What differentiated that object from other, similar NEAs, was its ephemeris 
which showed that it could be observed in favourable circumstances
($V<18\,\mathrm{mag}$) along a lengthy arc on the sky.  Its beginning was
obscured by the Milky Way, but from 2~March till 7~March the length of the
trail was $170\degr$, with the solar phase angle spanning the interval from
$36\degr$ to $86\degr$ (Fig.~\ref{PathInSky}).  Such observing geometry made
it possible to not only determine the rotation period and colour indices,
but also to attempt derivation of the magnitude-phase angle relation, spin
axis coordinates and shape.

We organized an observing campaign for 2021~DW$_1$ contacting a number of collaborating
observatories. A call for observations was also posted on the Minor Planet
Mailing List (MPML). As a result we set up a network of ten observatories
located on the Northern hemisphere, from South Korea in the East, to Arizona
in The West. The spread in longitude helped to secure observations from
different parts of the asteroid trail. There were several factors which made
observations difficult: the weather, bright Moon, fast movement of asteroid
in the sky, and human mistakes. We also
encountered strange problems with pointing some robotic telescopes. They
were probably caused by the systems using outdated ephemerides of
2021~DW$_{1}$. This once again proved a golden rule for this kind of
target of opportunity observations: to try to schedule observations, even if
any two of them were to be done simultaneously.

\begin{figure*}
\centering
    \begin{subfigure}{.42\textwidth}
    \includegraphics[width=\linewidth]{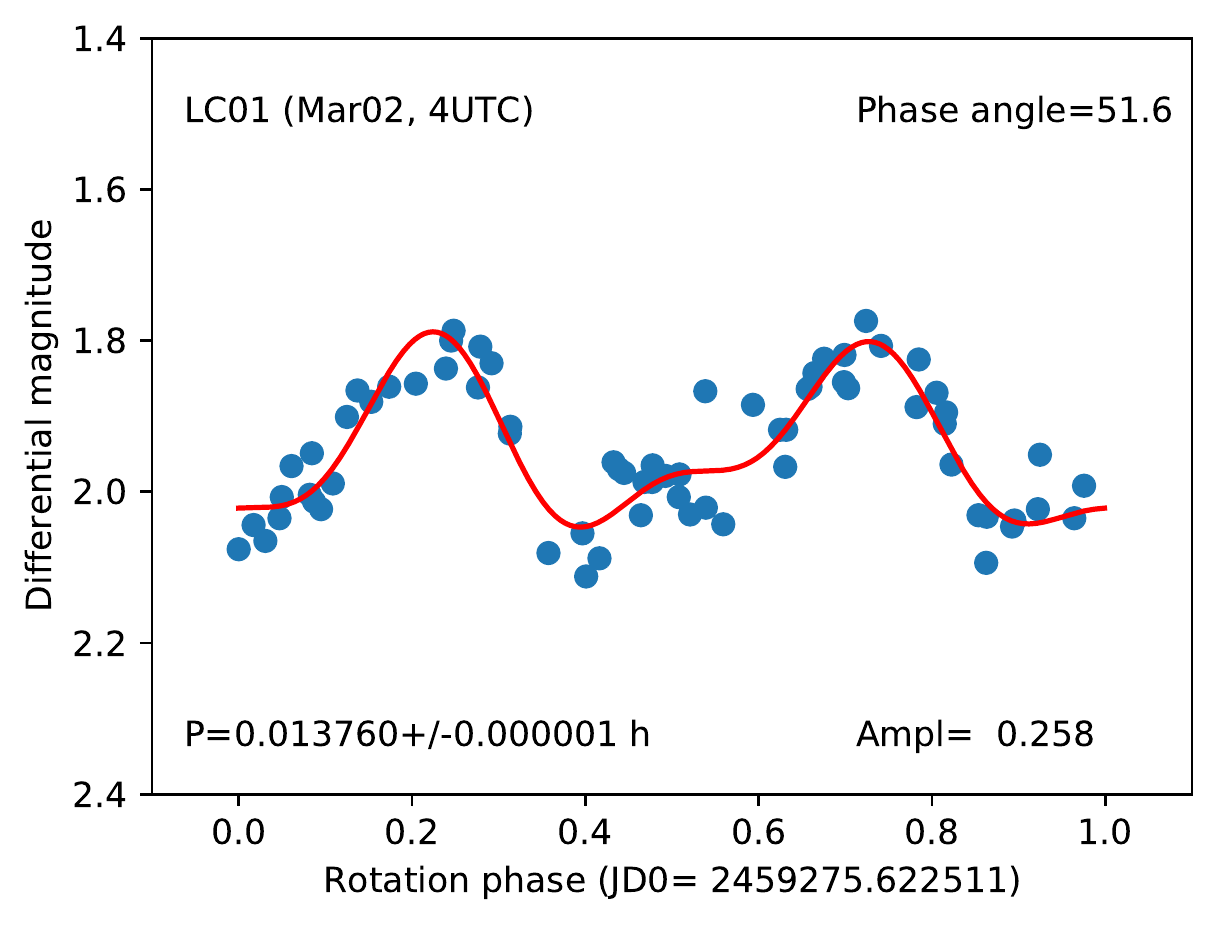}
    \end{subfigure}%
    \begin{subfigure}{.42\textwidth}
    \includegraphics[width=\linewidth]{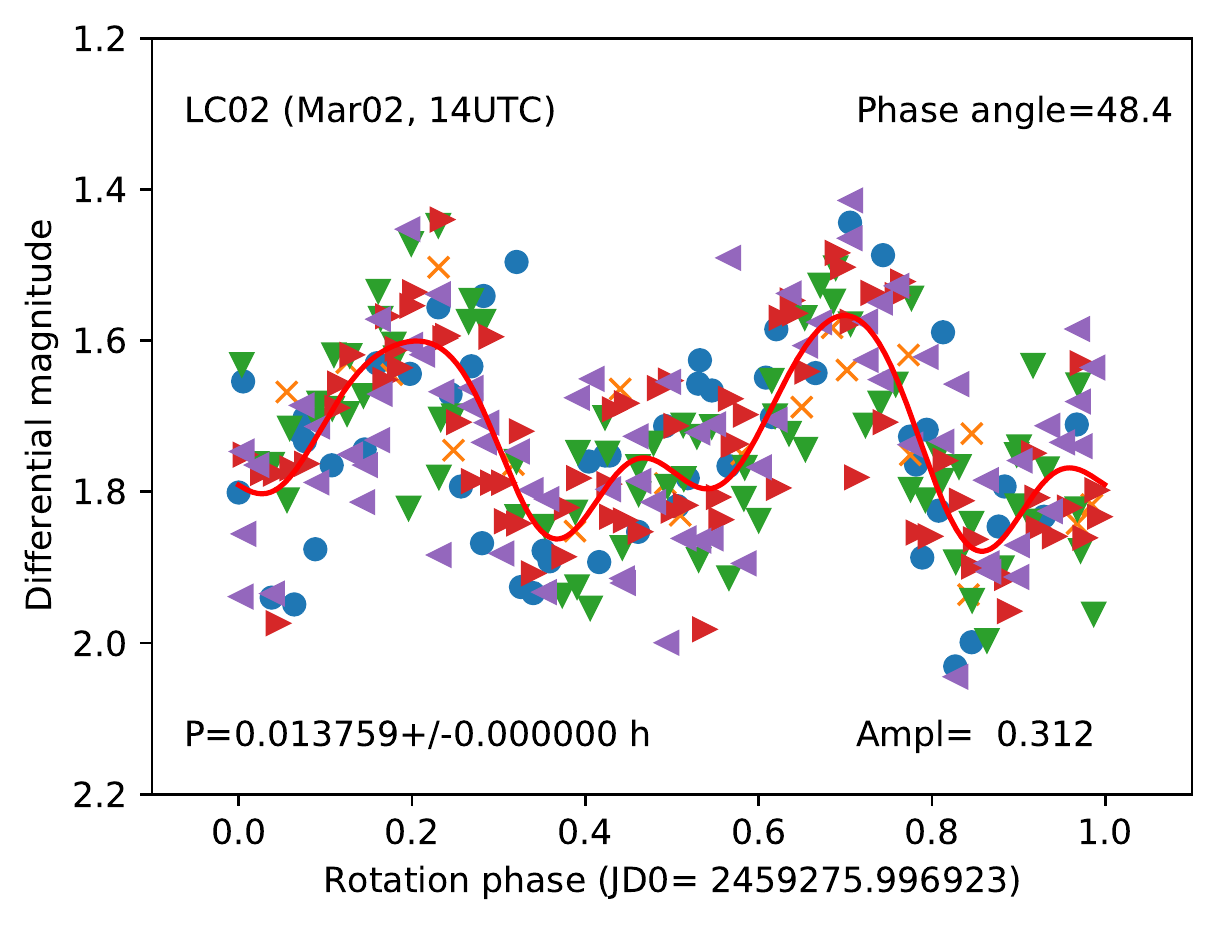}
    \end{subfigure}%
    \hfill
    \begin{subfigure}{.42\textwidth}
    \includegraphics[width=\linewidth]{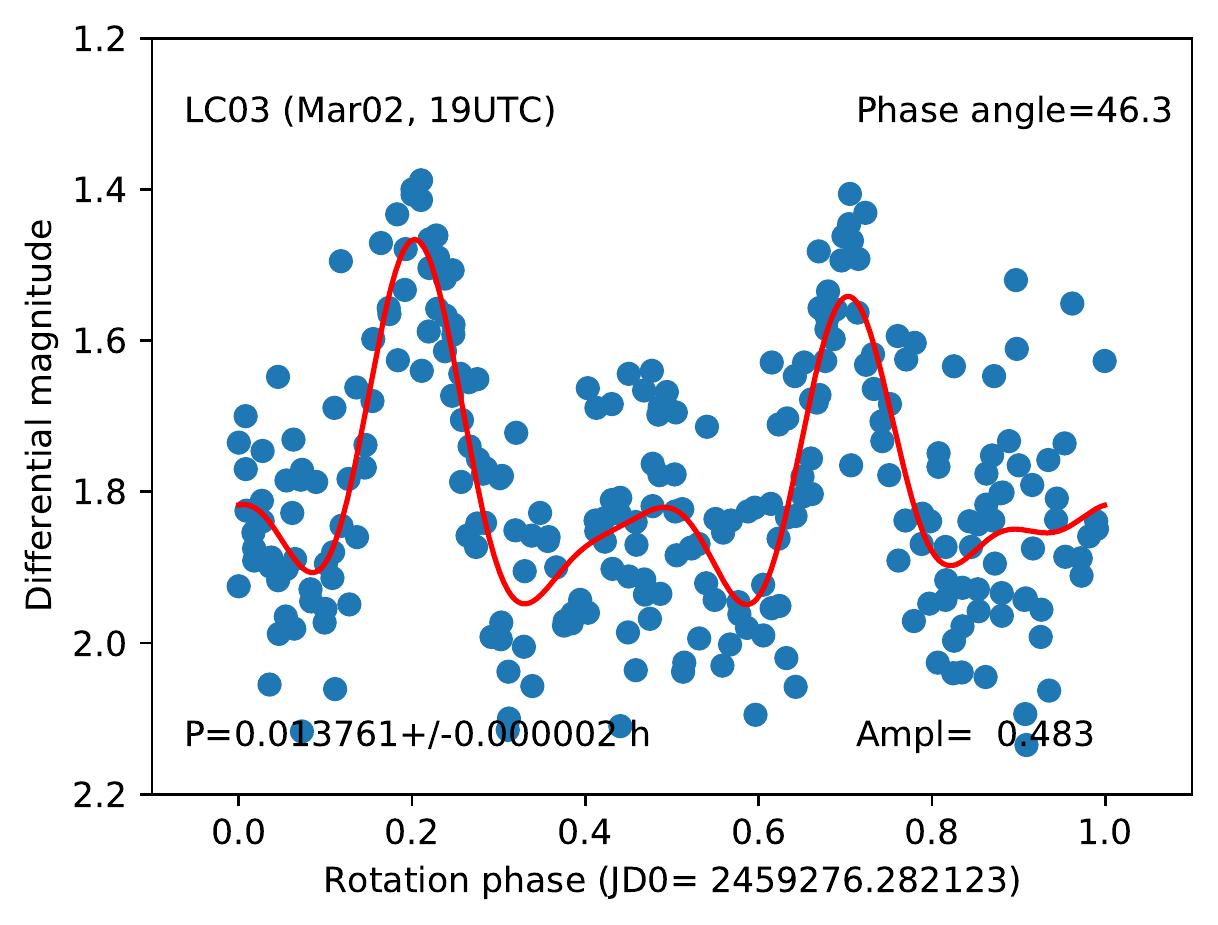}
    \end{subfigure}%
    \begin{subfigure}{.42\textwidth}
    \includegraphics[width=\linewidth]{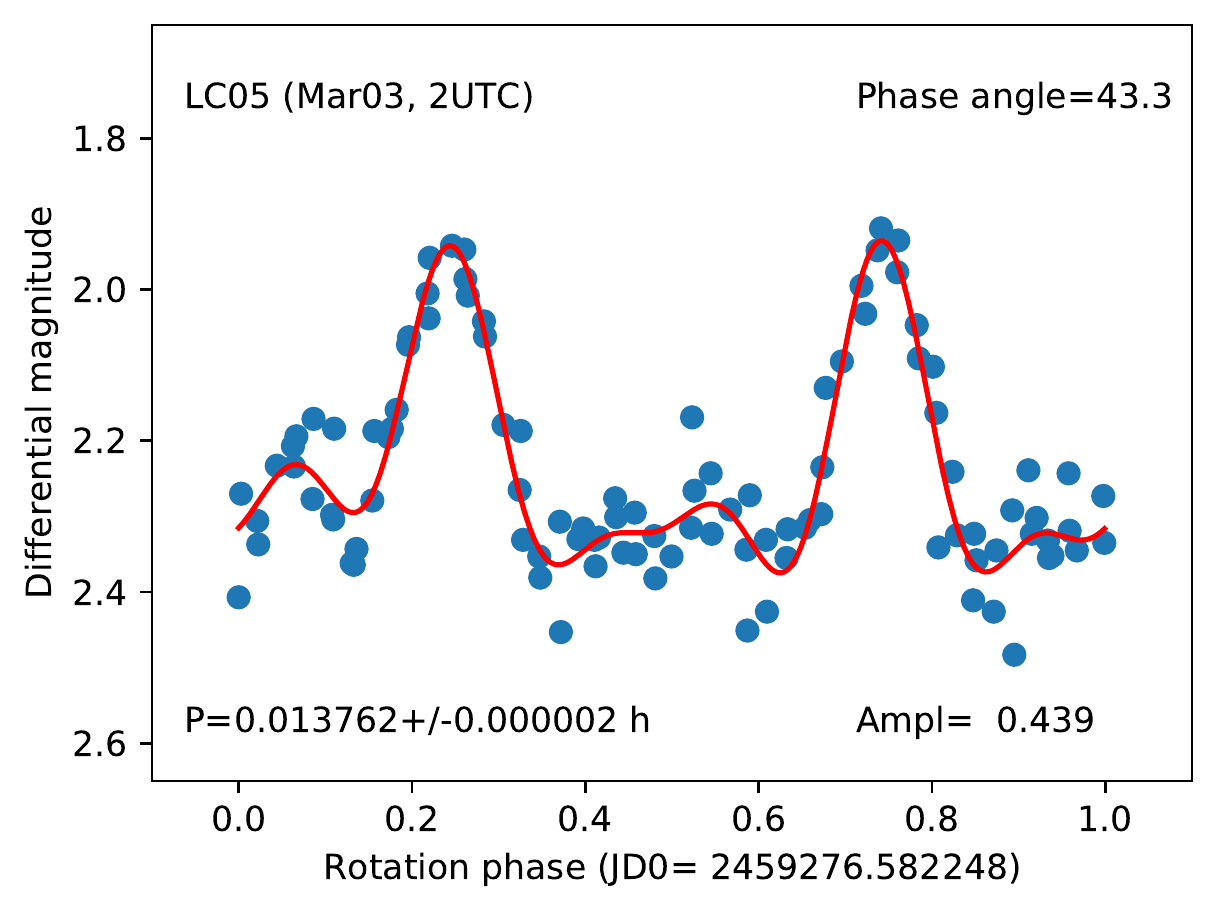}
    \end{subfigure}%
    \hfill
    \begin{subfigure}{.42\textwidth}
    \includegraphics[width=\linewidth]{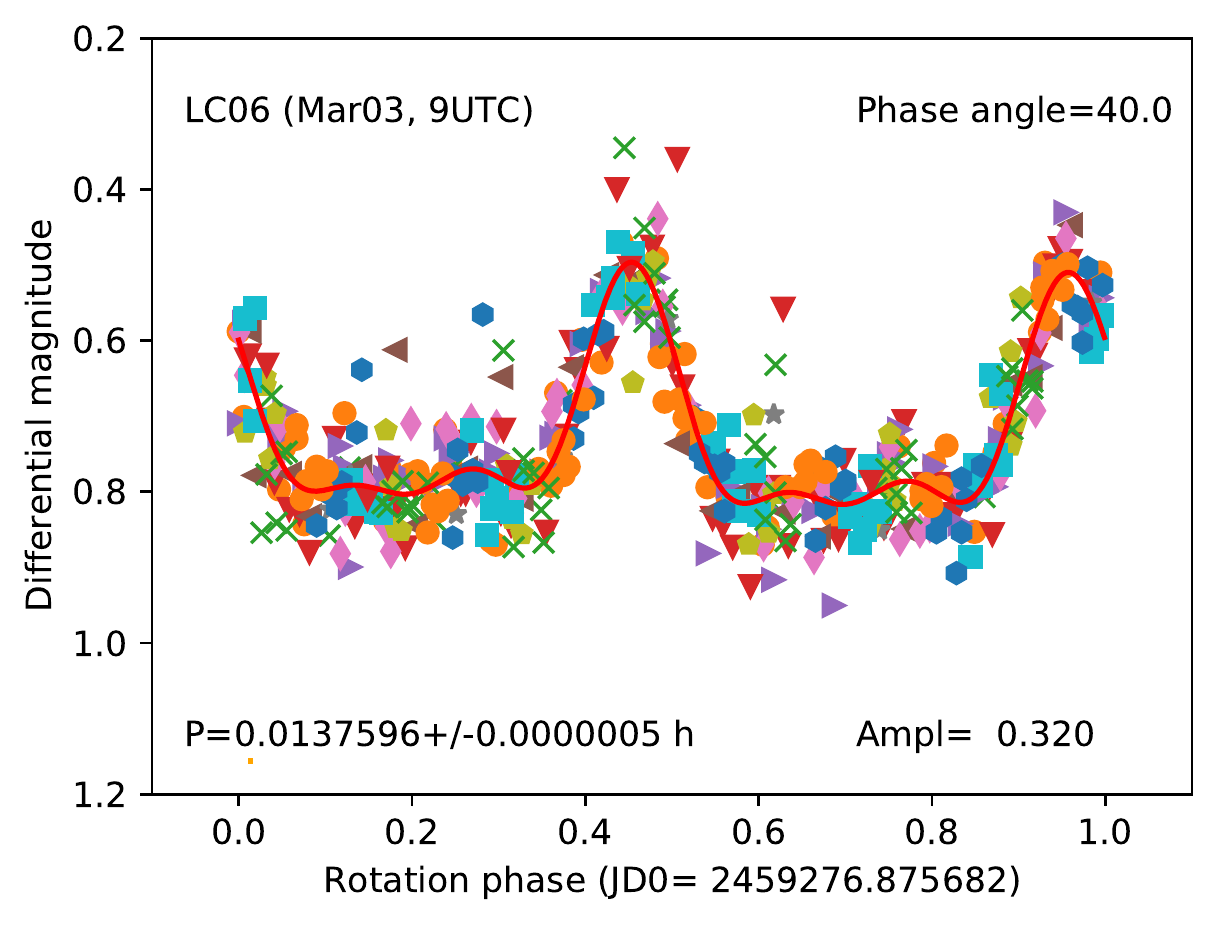}
    \end{subfigure}%
    \begin{subfigure}{.42\textwidth}
    \includegraphics[width=\linewidth]{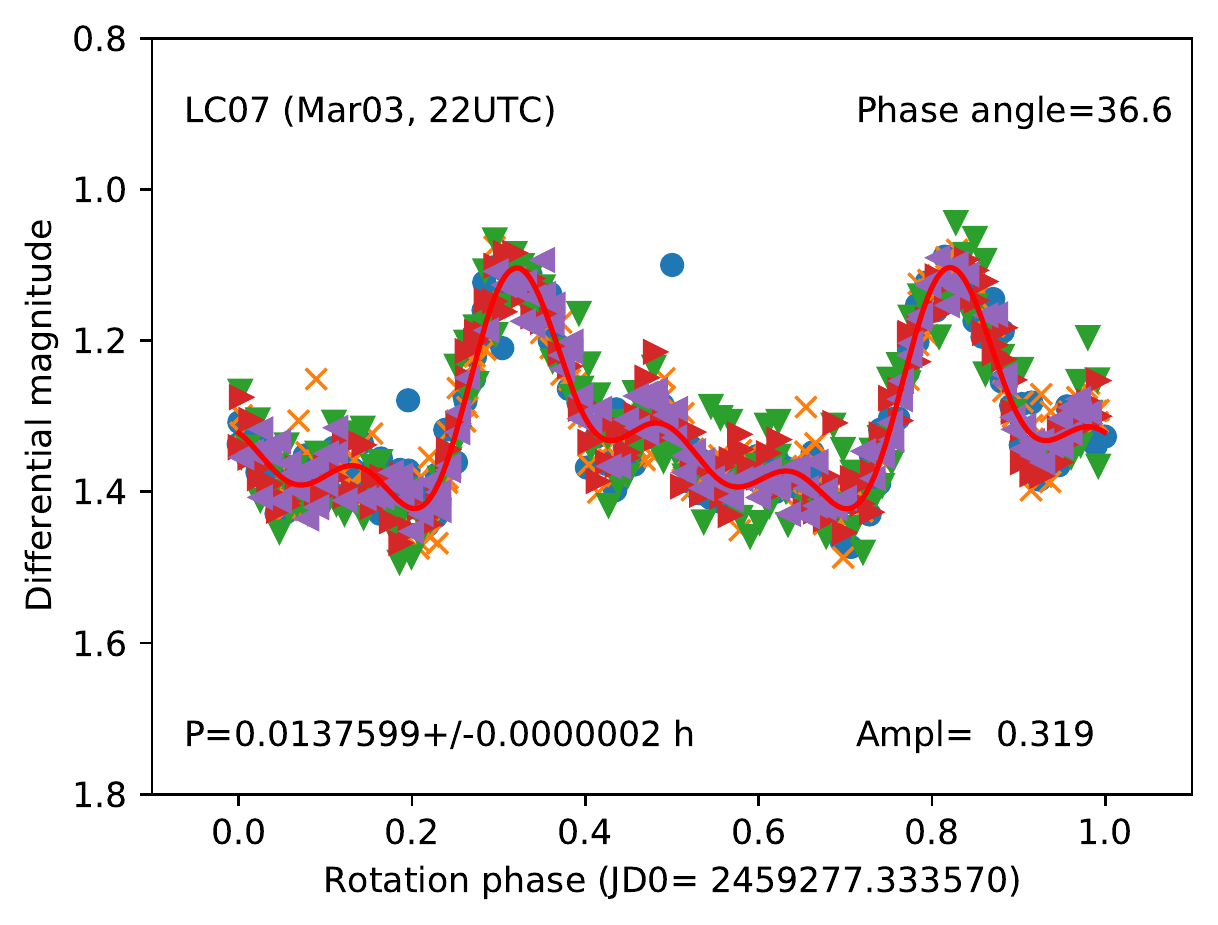}
    \end{subfigure}%
    \hfill
    \begin{subfigure}{.42\textwidth}
    \includegraphics[width=\linewidth]{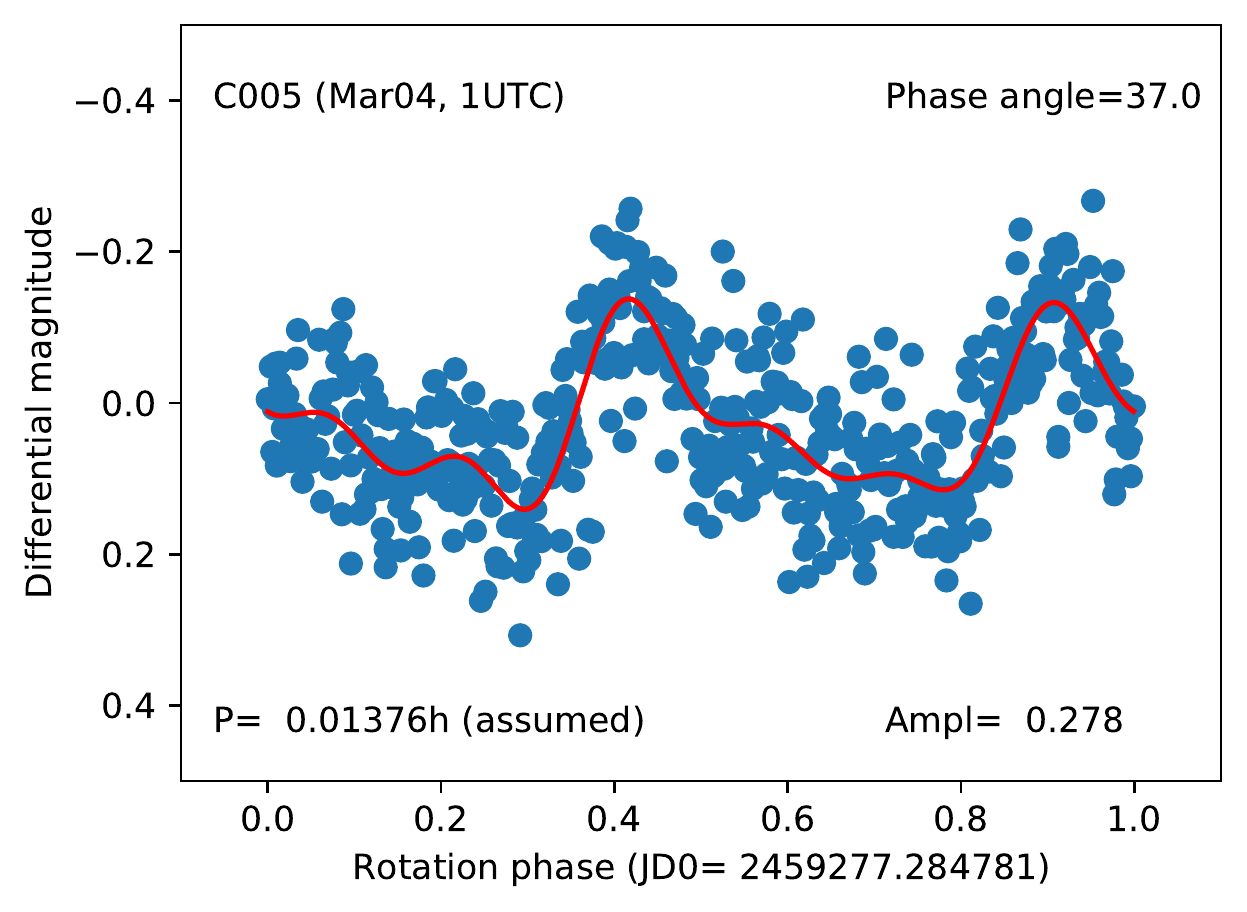}
    \end{subfigure}%
    \begin{subfigure}{.42\textwidth}
    \includegraphics[width=\linewidth]{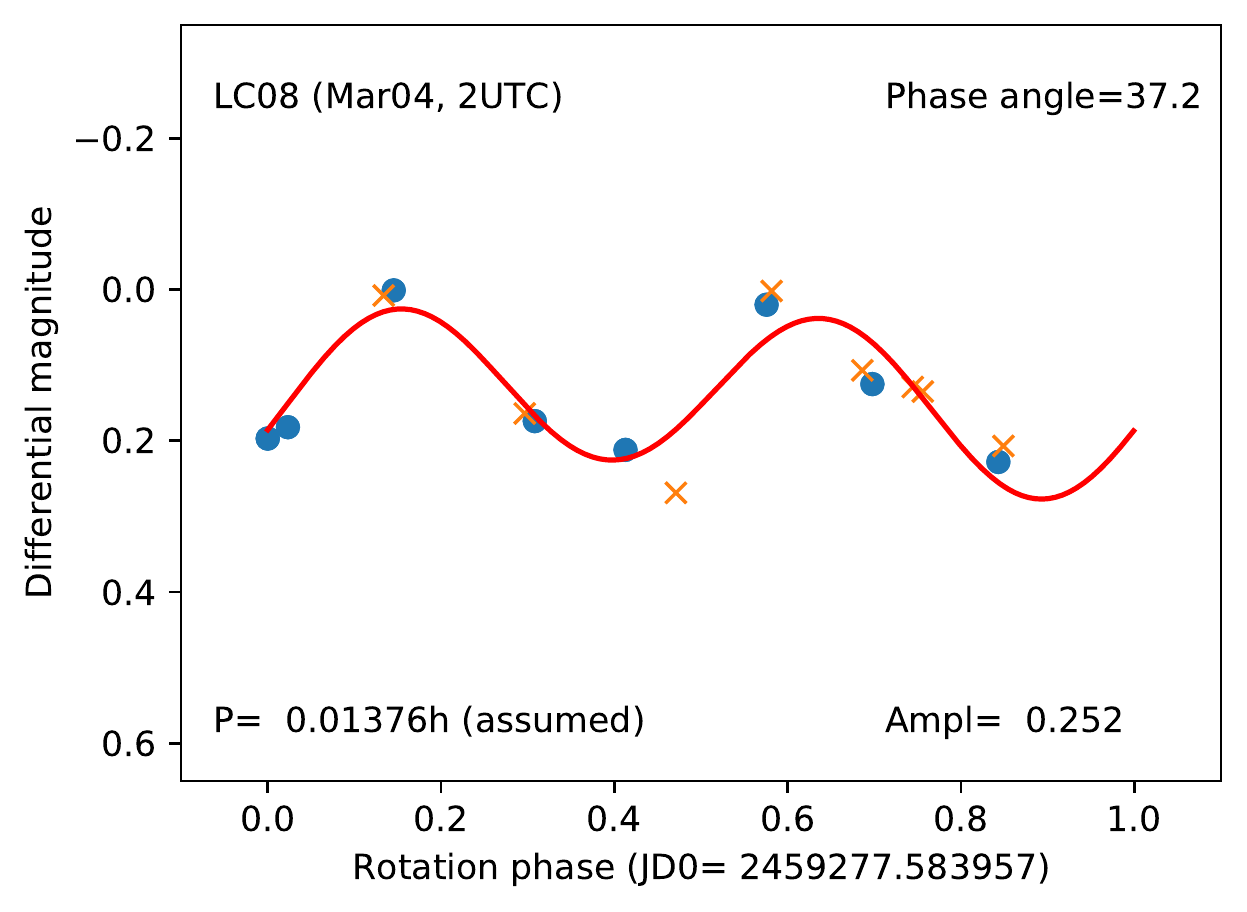}
    \end{subfigure}%
    \hfill    
\caption{Selected composite lightcurves of 2021~DW$_{1}$}
\label{LCs1}
\end{figure*}

The asteroid aspect data and the observing log are presented in
Table~\ref{AspectData}. We limit this table only to those observations, that
were used in final analysis (their results are shown in
Fig.~\ref{LCs1}). Details of all observing systems are shown in
Table~\ref{PhotSys}. 

Most of the photometric data were reduced with the Starlink
package\footnote{The Starlink software is currently supported by the East
Asian Observatory.}\citep{Currie+2013}.  Raw CCD frames were corrected for
bias and flat-field (and for the dark current, if necessary).  The aperture
photometry was then performed on the frames, on which the PSF (Point Spread
Function) of the asteroid and the comparison stars was almost circular.  On
the nights when fast sky motion of the object (even during short, 5~s
exposures) caused significant PSF trailing, we requested observers to employ
the non-sidereal telescope tracking on the asteroid.  This way, all signal
from 2021~DW$_{1}$ was concentrated in a circular aperture, while the images
of stars were trailed.  To perform differential photometry on such CCD
frames, we used ''pill-shaped apertures'' developed by \citet{Fraser+2016}, 
and used for NEAs by \citet{Kolenczuk2020}.  In this
technique, an elongated aperture is created as a rectangle with two
semicircular end-caps.  It is defined by three parameters: the trail length,
the position angle, and the radius.

The data from the Great Shefford Observatory were reduced in a standard way
with a help of Astrometrica and MPO~Canopus commercial packages. 
Observations from the Platanus Observatory were done with an exposure time
of 1~s.  This produced a circular PSF for a fast moving asteroid, but because
of a small telescope aperture (0.28-m), the SNR was low.  However, during a
one-hour run, more than two thousand frames were collected.  After reducing
the data, every five points were averaged into one data point, and
the composite lightcurve, even if still quite noisy, produced a well-defined
Fourier series fit (see the C005 plot on Fig.~\ref{LCs1}).

\begin{table*}
\caption{Observing log.}
\label{AspectData}
\begin{center}
\begin{tabular}{cclccrrrrrrcrrr}
\hline \hline
\multicolumn{1}{c}{Date     } &
\multicolumn{1}{c}{Obs. time} &
\multicolumn{1}{c}{Observatory} &
\multicolumn{1}{c}{r}         &
\multicolumn{1}{c}{$\Delta$}  &
\multicolumn{1}{c}{$\alpha$}     &
\multicolumn{1}{c}{$\lambda$} &
\multicolumn{1}{c}{$\beta$}   &
\multicolumn{1}{c}{$V$}         &
\multicolumn{1}{c}{Mov}       &
\multicolumn{1}{c}{Exp}       &
\multicolumn{1}{c}{Fltr}      &
\multicolumn{1}{c}{Code} \\
\multicolumn{1}{c}{(UTC)}     &
\multicolumn{1}{c}{(UTC)}     &
\multicolumn{1}{c}{(see Table~\ref{PhotSys})} &
\multicolumn{1}{c}{(au)}      &
\multicolumn{1}{c}{(au)}      &
\multicolumn{1}{c}{$(\degr)$} &
\multicolumn{1}{c}{$(\degr)$} &
\multicolumn{1}{c}{$(\degr)$} &
\multicolumn{1}{c}{[mag]}     &
\multicolumn{1}{c}{''/min}    &
\multicolumn{1}{c}{[s]}       &
                              &
                              \\
\hline
2021-03-02 & 02:56 -- 03:36 &  Lowell (1)   & 0.9959 & 0.0078 & 51.6 & 114.6 & -25.0 & 16.4 & 25  & 10  & VR & LC01\\
2021-03-02 & 11:55 -- 15:11 &  DOAO (2)     & 0.9956 & 0.0068 & 48.4 & 116.7 & -19.9 & 16.0 & 35  & 10  &  I & LC02 \\
2021-03-02 & 18:46 -- 19:35 &  Schiap. (3)  & 0.9955 & 0.0062 & 46.2 & 118.2 & -16.2 & 15.7 & 43  &  5  &  C & LC03 \\
2021-03-03 & 01:58 -- 02:22 &  Lowell (1)   & 0.9953 & 0.0055 & 43.2 & 120.3 & -10.7 & 15.4 & 55  &  4  & VR & LC05 \\
2021-03-03 & 09:01 -- 09:59 &  Winer (5)    & 0.9951 & 0.0049 & 39.9 & 122.9 & -03.4 & 15.0 & 69  &  5  &  C & LC06 \\
2021-03-03 & 20:00 -- 21:49 &  Mayaki (6)   & 0.9948 & 0.0041 & 36.5 & 128.6 & +12.2 & 14.5 & 98  &  3  &  C & LC07 \\
2021-03-04 & 01:07 -- 02:07 &  Platanus (10)& 0.9946 & 0.0039 & 37.0 & 131.6 & +19.9 & 14.4 &112  &  $1^{*}$  &  C & C005 \\
2021-03-04 & 02:00 -- 02:12 &  McDonald (7) & 0.9946 & 0.0039 & 37.1 & 131.9 & +20.6 & 14.4 &112  &  5  &  w & LC08 \\
2021-03-04 & 19:22 -- 20:26 &  DOAO (2)     & 0.9942 & 0.0041 & 52.4 & 150.8 & +51.3 & 15.0 & 98  &  5  &  I & LC09 \\
2021-03-05 & 02:08 -- 02:21 &  CalarAlto (8)& 0.9940 & 0.0044 & 59.0 & 161.0 & +59.1 & 15.3 & 88  &2.5  &  C & LC10 \\
2021-03-05 & 03:30 -- 03:59 &  Lowell (1)   & 0.9940 & 0.0045 & 60.5 & 164.0 & +60.8 & 15.5 & 84  &  4  & VR & A001 \\
2021-03-05 & 06:00 -- 06:44 &  Lowell (1)   & 0.9939 & 0.0047 & 63.1 & 169.6 & +63.3 & 15.8 & 77  &  4  & VR & A008 \\
2021-03-05 & 06:45 -- 06:51 &  Lowell (1)   & 0.9939 & 0.0047 & 63.5 & 170.4 & +63.6 & 15.8 & 76  &  4  &  g & CI01 \\ 
2021-03-05 & 06:52 -- 06:58 &  Lowell (1)   & 0.9939 & 0.0047 & 63.6 & 170.6 & +63.7 & 15.8 & 76  &  4  &  i & CI02 \\
2021-03-05 & 06:59 -- 07:05 &  Lowell (1)   & 0.9939 & 0.0047 & 63.7 & 170.8 & +63.8 & 15.8 & 76  &  4  &  z & CI03 \\
2021-03-05 & 11:40 -- 12.17 &  Winer (5)    & 0.9938 & 0.0052 & 68.0 & 182.4 & +67.1 & 15.9 & 59  &  5  &  C & A016 \\
2021-03-06 & 00:37 -- 02:32 &  Shefford (9) & 0.9935 & 0.0064 & 76.7 & 214.1 & +70.3 & 16.7 & 38  &  4  &  C & LC12 \\ 
2021-03-06 & 08:34 -- 10:46 &  Winer (5)    & 0.9934 & 0.0073 & 80.1 & 227.8 & +69.7 & 17.1 & 28  &  5  &  C & B579 \\
2021-03-07 & 08:45 -- 08:52 &  McDonald (7) & 0.9929 & 0.0100 & 86.6 & 250.1 & +66.2 & 18.2 & 17  &  4  &  w & LC13 \\
\hline \hline
\end{tabular}
\end{center}
Note: the table presents the subset of the observing log limited to the lightcurves
which were used in the analysis. The third column shows the shortened observatory name (full names are presented in
Table~\ref{PhotSys}). The next five columns present the aspect data for the middle of the
observing time: $r$ and $\Delta$ are the distances of the asteroid from the Sun and the Earth, respectively, 
$\alpha$ is the solar phase angle, while $\lambda$ and $\beta$ are the geocentric, ecliptic (J2000) longitude and latitude.
In the next column, an average brightness $V$ of the asteroid, as predicted by the Horizons ephemeris, is given. 
Starting from the tenth column, the table gives the asteroid movement on the sky (Mov), the exposure time (Exp), and the 
filter used in the observations (here "C" stands for a "clear" filter). The last column provides the code to locate the lightcurve in Fig.~\ref{LCs1}. \\
(*) For C005 lightcurve more than 2000 exposures were obtained, with a very short exposure time of 1~s. They were then averaged (every five points into one) which gave satisfactory result.
\end{table*}

\begin{table*}
\caption{Observatories, telescopes and detectors used in observations}
\label{PhotSys}
\begin{center}
\begin{tabular}{lll}
\hline \hline
\multicolumn{1}{c}{Observatory} & 
\multicolumn{1}{c}{Telescope} &
\multicolumn{1}{c}{Detector} \\
\hline \hline
(1) Anderson-Mesa, Lowell Obs. (IAU 688), Arizona & 1.1-m Hall & e2v CCD231 \\
(2) Deokheong Optical Astronomy Obs. (IAU P66), South Korea & 1.0-m RC & PI SOPHIA-2048B CCD \\
(3) Schiaparelli Obs. (IAU 204), Italy & 0.8-m & SBIG STX-16803 CCD \\
(4) San Marcello Pistoiese Obs. (IAU 104), Italy & 0.6-m & CCD \\
(5) Winer Obs. (IAU 648), Arizona & 0.7-m RBT/PST2 & Andor iXon 888 CCD\\
(6) Odessa-Mayaki Obs. (IAU 583), Ukraine & 0.8-m OMT-800 & FLI ML09000 CCD\\
(7) McDonald Obs. LCO (IAU V37), Texas & 1.0-m 1m0-08 & FLI ML4720 CCD\\
(8) Calar Alto Obs. (IAU 493), Spain & 1.23-m & e2v CCD231-84 \\
(9) Great Shefford Obs. (IAU J95), Great Britain & 0.4-m &  Apogee Alta U47+ CCD \\
(10) Planatus Observatory (IAU K80), Poland & 0.28-m & ZWO ASI290MM CMOS \\
\hline\hline
\end{tabular}
\end{center}
Note: the table includes information about all observatories which provided
data on 2021~DW$_1$. Some of them were used only for cross-validation and
are not presented in this paper.
\end{table*}

To search for the rotation period of 2021~DW$_{1}$ we used a standard
Fourier series analysis.  This was done with a PerFit program
\citep{Kwiatkowski+2010}, which in its current version works 
in the following way: after correcting all times for the light-time,
each differential photometry time series is divided into parts obtained with
the same comparison star.  Next, a Fourier series of a given order is fit
to all of them with one synodic period assumed.  In the process, both the
Fourier series coefficients and magnitude shifts (with respect to the first
time series) are determined by linear least squares.  This process is
repeated for each trial period from a specified interval.  In most cases the
4th to 6th order Fourier series is used. The accuracy of the fit is measured
with the $\chi^2$ computed from the residuals.

The period which gives the smallest $\chi^2$ 
is then selected as a potential
synodic period of rotation. The least square fit takes into account the
accuracy of the measured asteroid differential magnitudes (expressed as
standard deviations). The accuracy of the
derived period is estimated by a Monte Carlo method, in which all data are
perturbed 30 times with the appropriate sigma values. 

\begin{figure*}
\begin{center}
    \begin{subfigure}{.42\textwidth}
    \includegraphics[width=\linewidth]{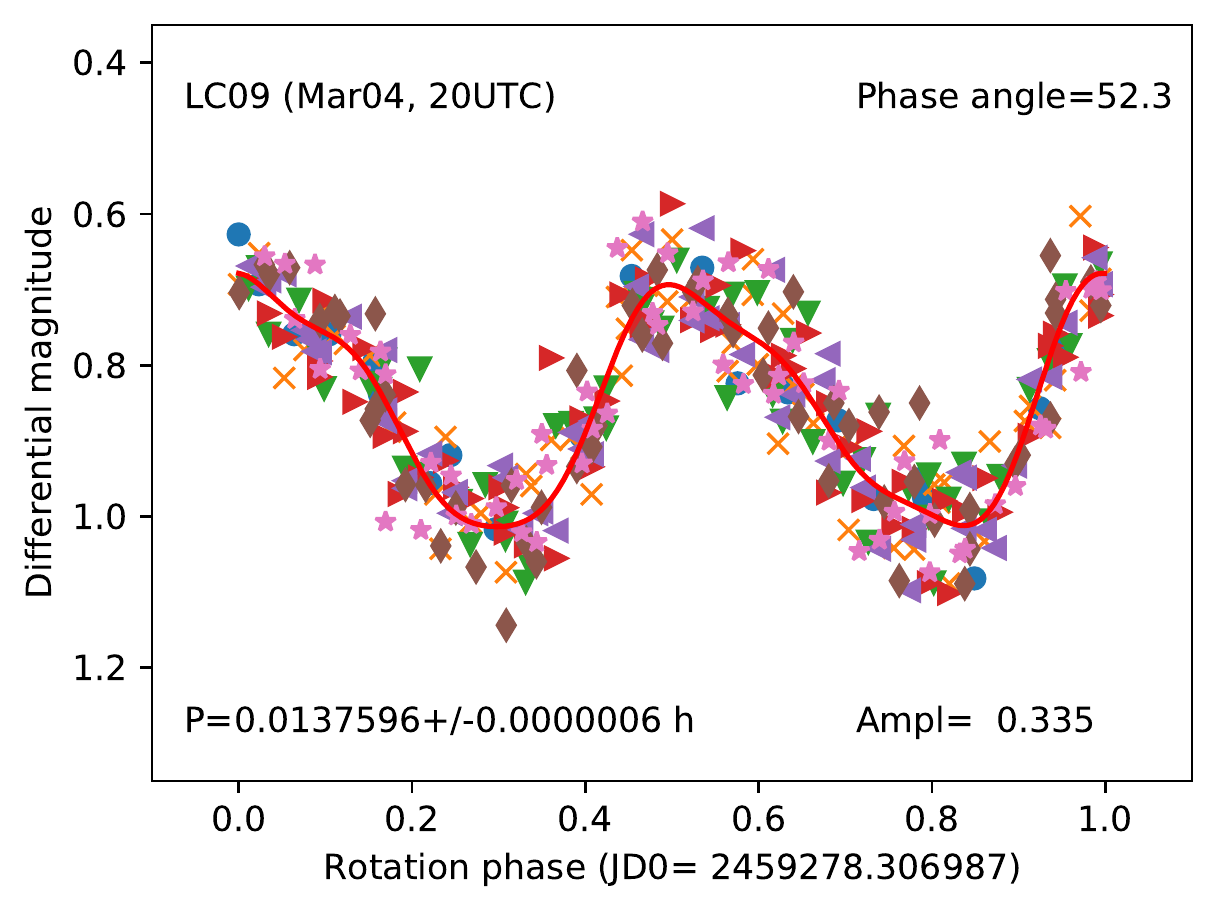}
    \end{subfigure}%
    \begin{subfigure}{.42\textwidth}
    \includegraphics[width=\linewidth]{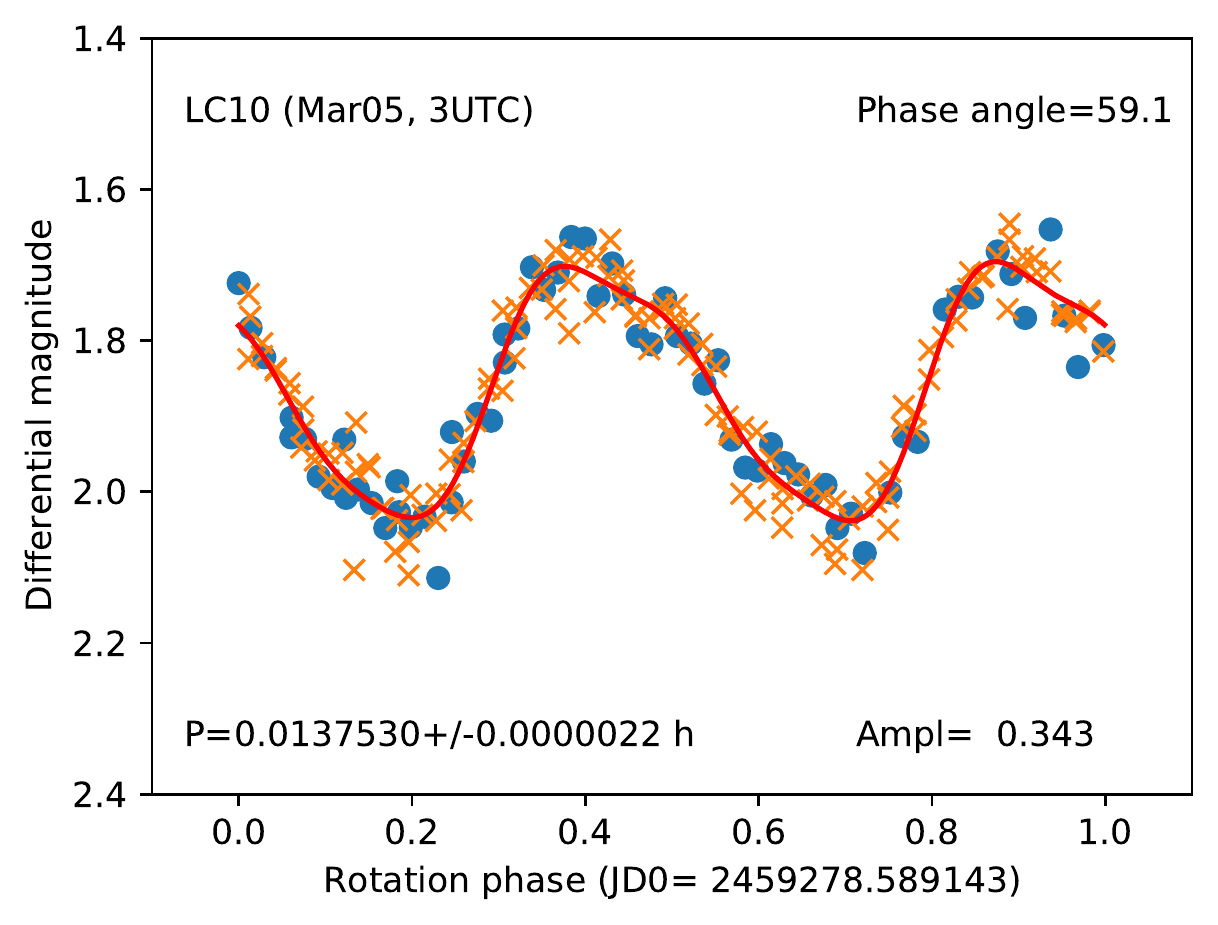}
    \end{subfigure}%
    \hfill
    \begin{subfigure}{.42\textwidth}
    \includegraphics[width=\linewidth]{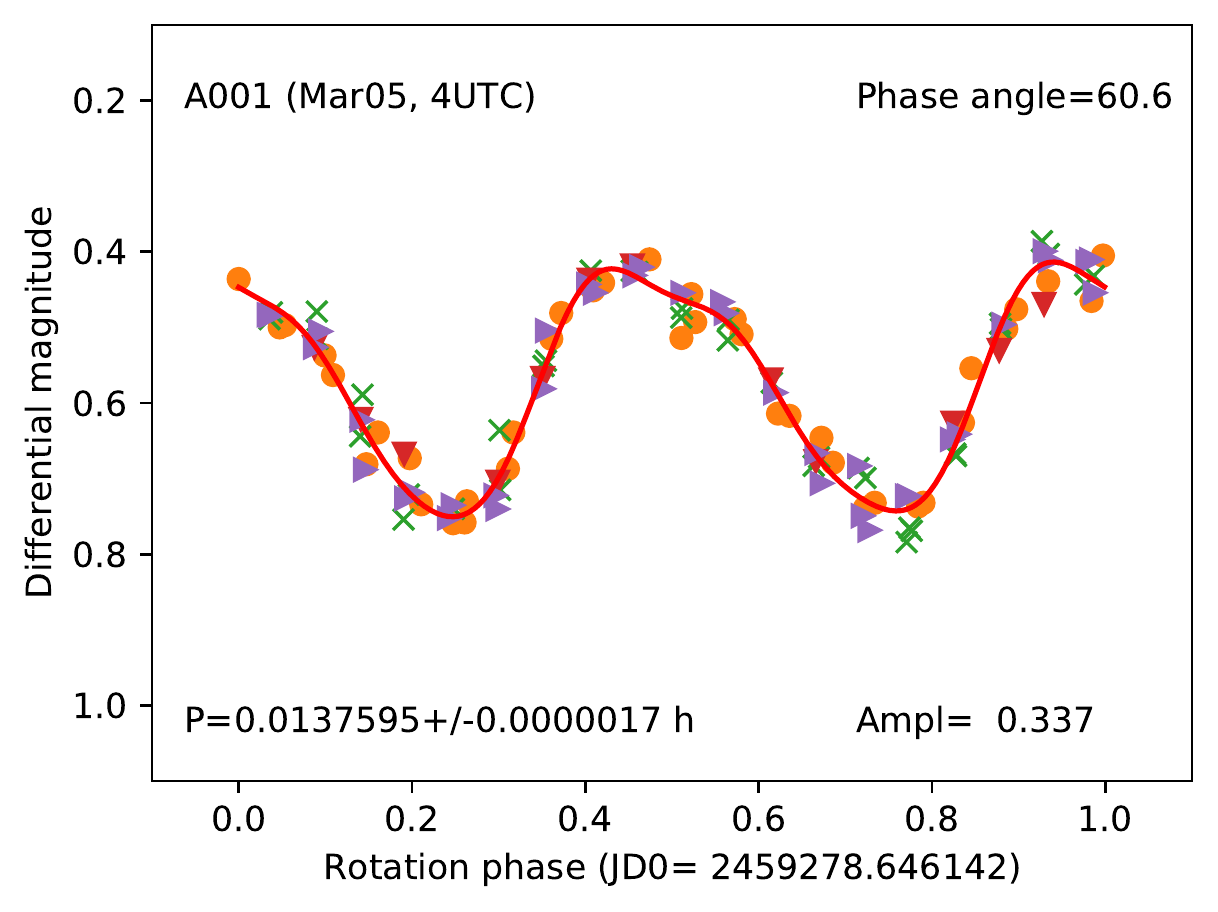}
    \end{subfigure}%
    \begin{subfigure}{.42\textwidth}
    \includegraphics[width=\linewidth]{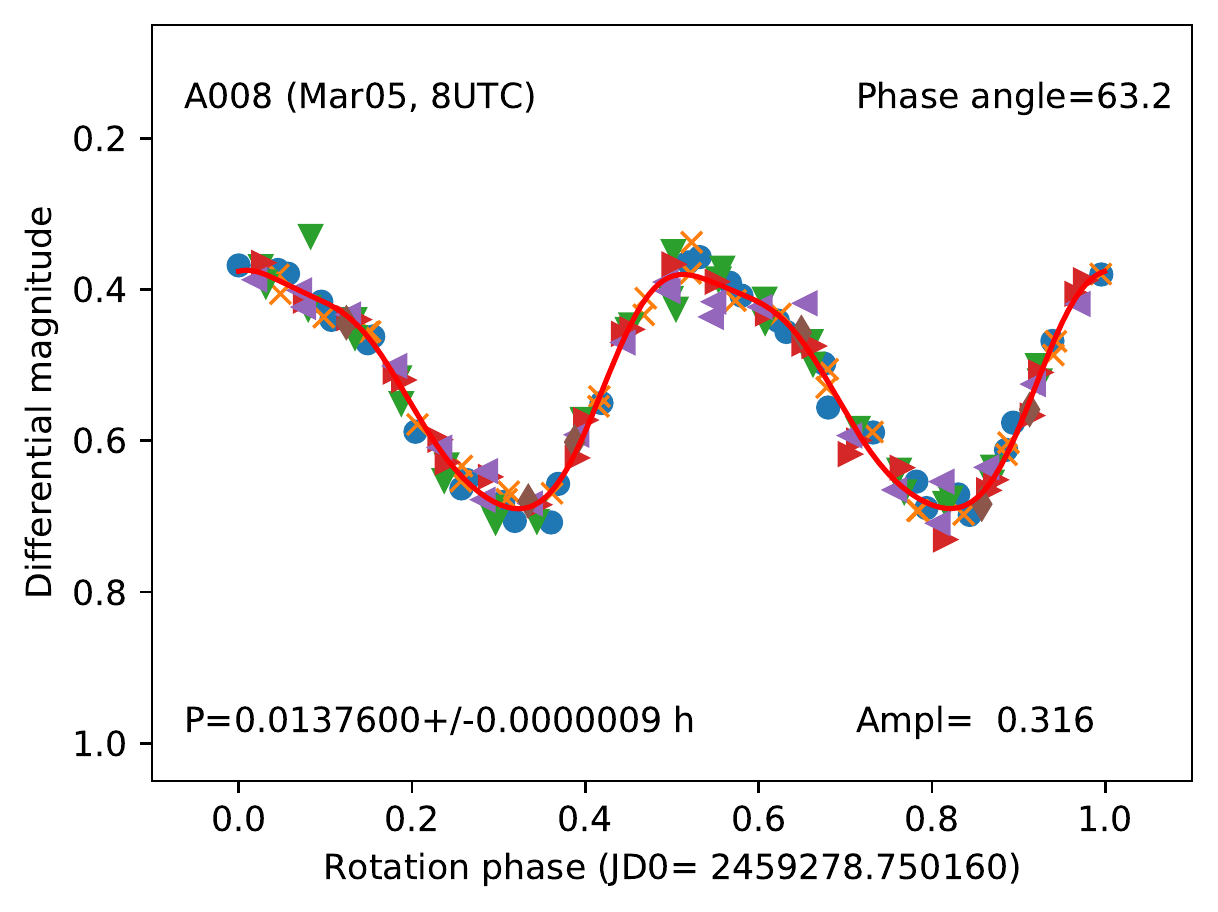}
    \end{subfigure}%
    \hfill
    \begin{subfigure}{.42\textwidth}
    \includegraphics[width=\linewidth]{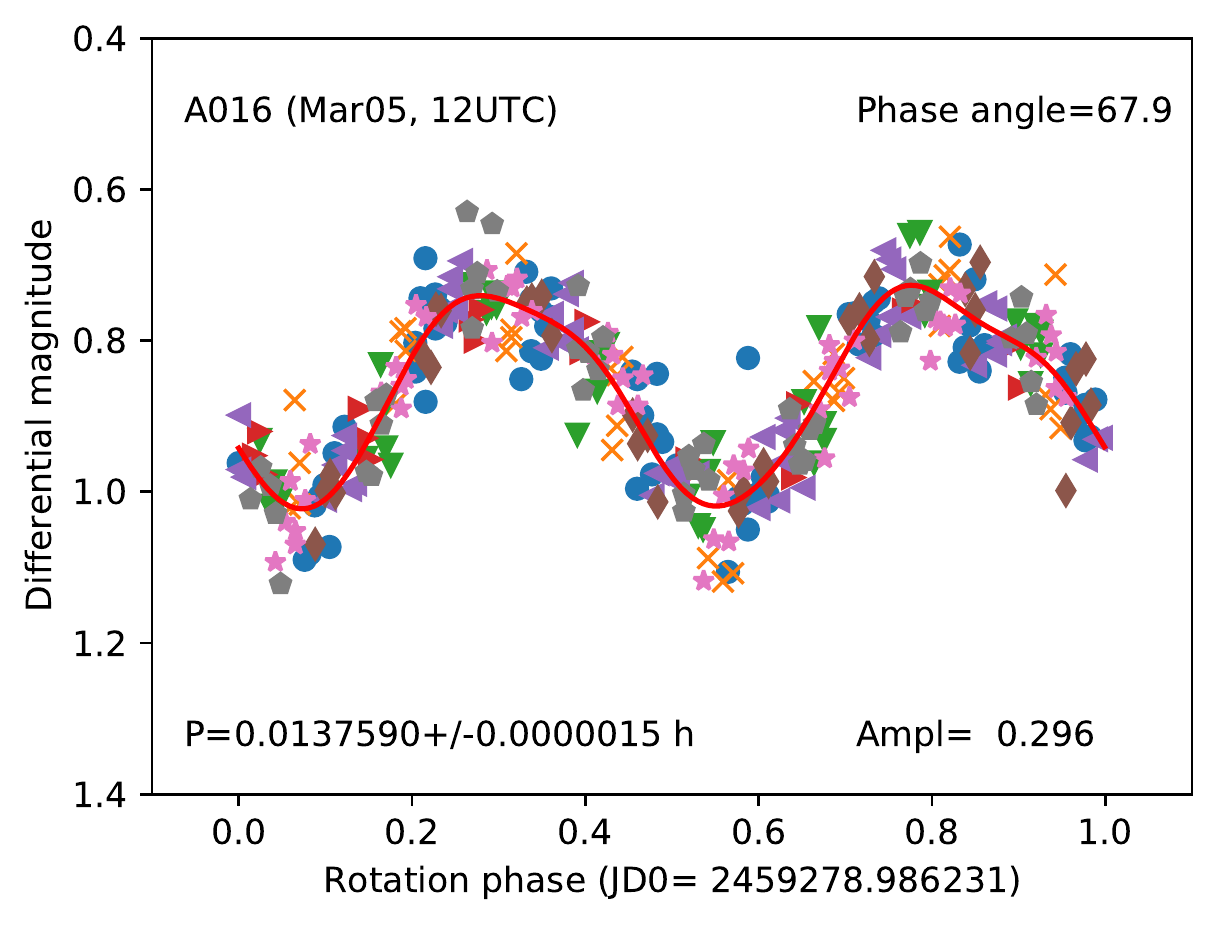}
    \end{subfigure}%
    \begin{subfigure}{.42\textwidth}
    \includegraphics[width=\linewidth]{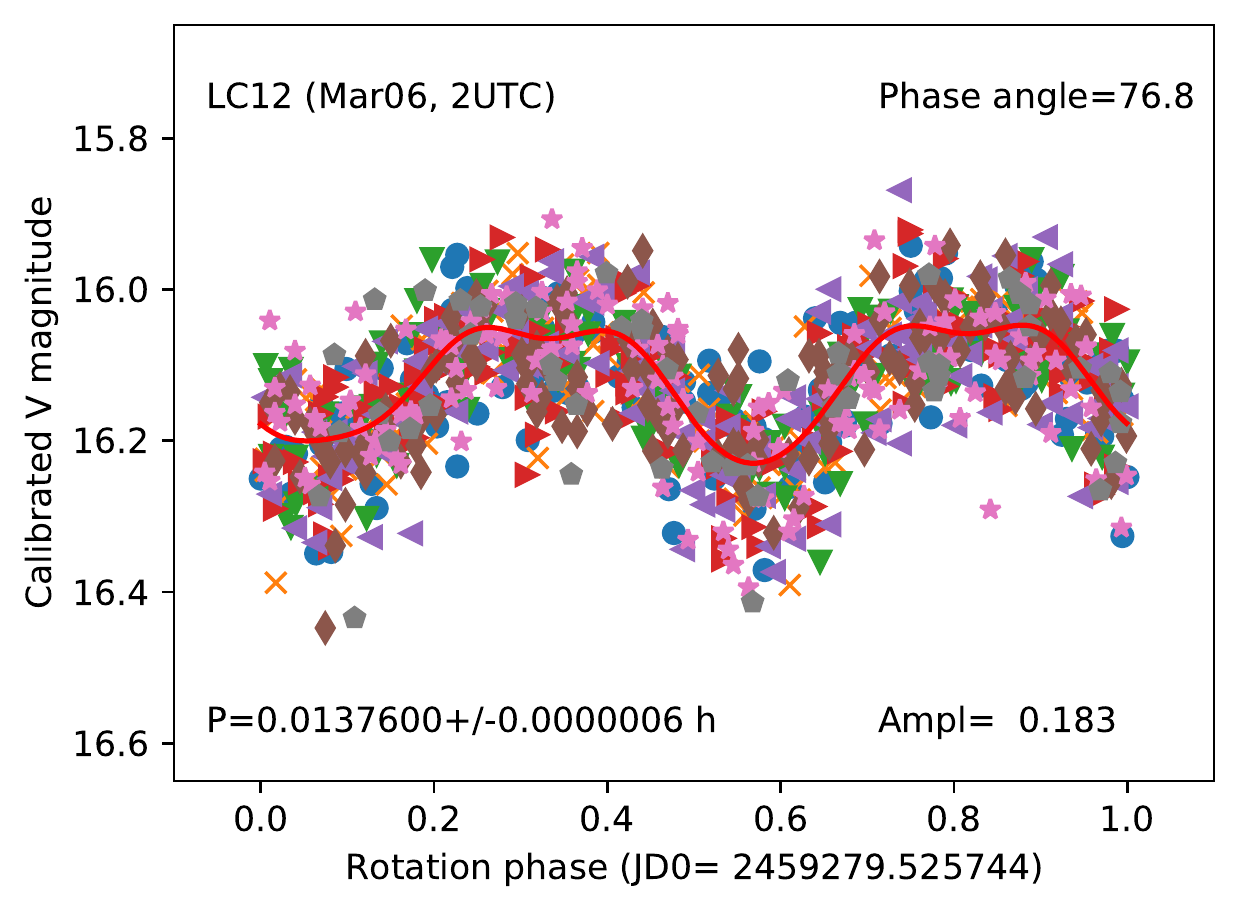}
    \end{subfigure}%
    \hfill
    \begin{subfigure}{.42\textwidth}
    \hspace{1em} \includegraphics[width=\linewidth]{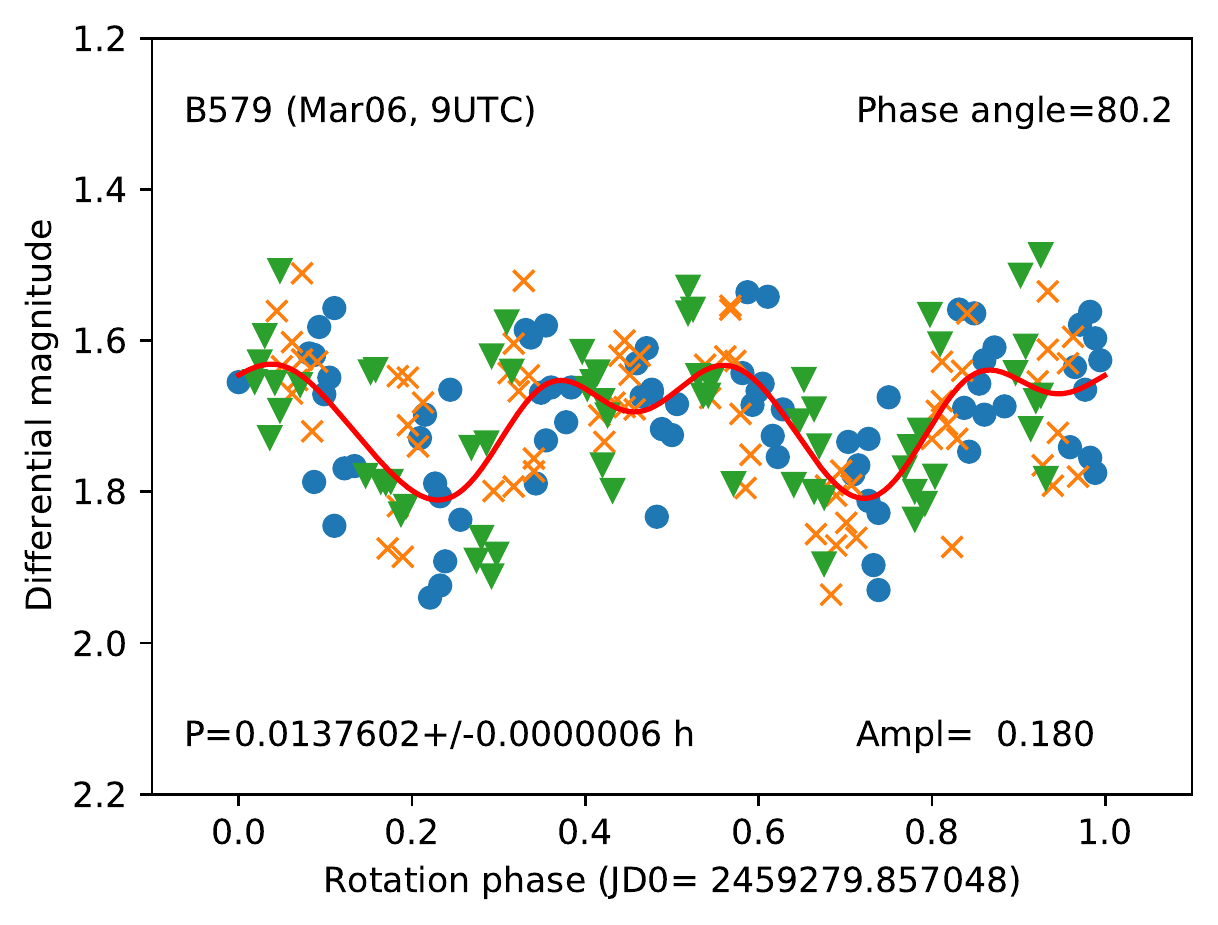}
    \end{subfigure}%
    \hfill
\end{center}

\begin{center}
    Fig.~2, cont.
\end{center}
\label{LCs2}
\end{figure*}

Using PerFit for the lightcurves of 2021~DW1 it appeared
there are two solutions for the period: $P_1=0.00688\,\mathrm{h}$ (this gave
mono-modal composite lightcurves) and $P_2=0.01376\,\mathrm{h}$ (bi-modal
composite lightcurves), with $P_2=2\cdot P_1$.  Results for
$P_2=0.01376\,\mathrm{h}$ (which is our preferred solution) are presented in
Fig.~\ref{LCs1}.  In the plots we do not draw error bars not to
obscure images.  The overall accuracy can be easily judged from the scatter
of the data points about the Fourier fit. 

\section{Sidereal period, spin axis and shape}

Since the works of \citet{Kaasalainen+2001} and \citet{Kaasalainen+2001a} 
a standard method for determination of the asteroid sidereal rotation
period, pole and shape is through the lightcurve inversion.  For that,
lightcurves observed at different viewing and illuminating geometries are
needed. While each case is different, a good rule of thumb for NEAs is to
have data sampling an arc on the sky longer than $120\degr$ (Josef \v{D}urech,
personal communication). While in case of 2021~DW$_1$ we observed its trail
extending up to $170\degr$, the last two observations (from 7 March)
were very noisy (the asteroid brightness dropped to $V=18\,\mathrm{mag}$)
and were not used in the analysis. This shortened the arc to $163\degr$,
from which we selected thirteen composite lighcurves well positioned along 
the trail (see Fig.~\ref{LCs1}). Each of them was obtained with
the $P_2$ synodical period. Their brightnesses
were converted from magnitudes to fluxes and rescaled so that the average 
flux during the rotation was set to unity.  We did not use any lightcurves
calibrated to the standard magnitudes, as such callibrations are usually less
accurate than the differential magnitudes.  The data were then used for the
lightcurve inversion, for which we used a C-language implementation of the
code, written by Josef \v{D}urech.  The convex shapes used in computations were
approximated by spherical harmonics of the 6th degree and order.  The light
scattering law was approximated by the Lommel-Seeliger-Lambert function,
with c=0.1.

\begin{figure}[!t]
\resizebox{\hsize}{!}{\includegraphics[clip]{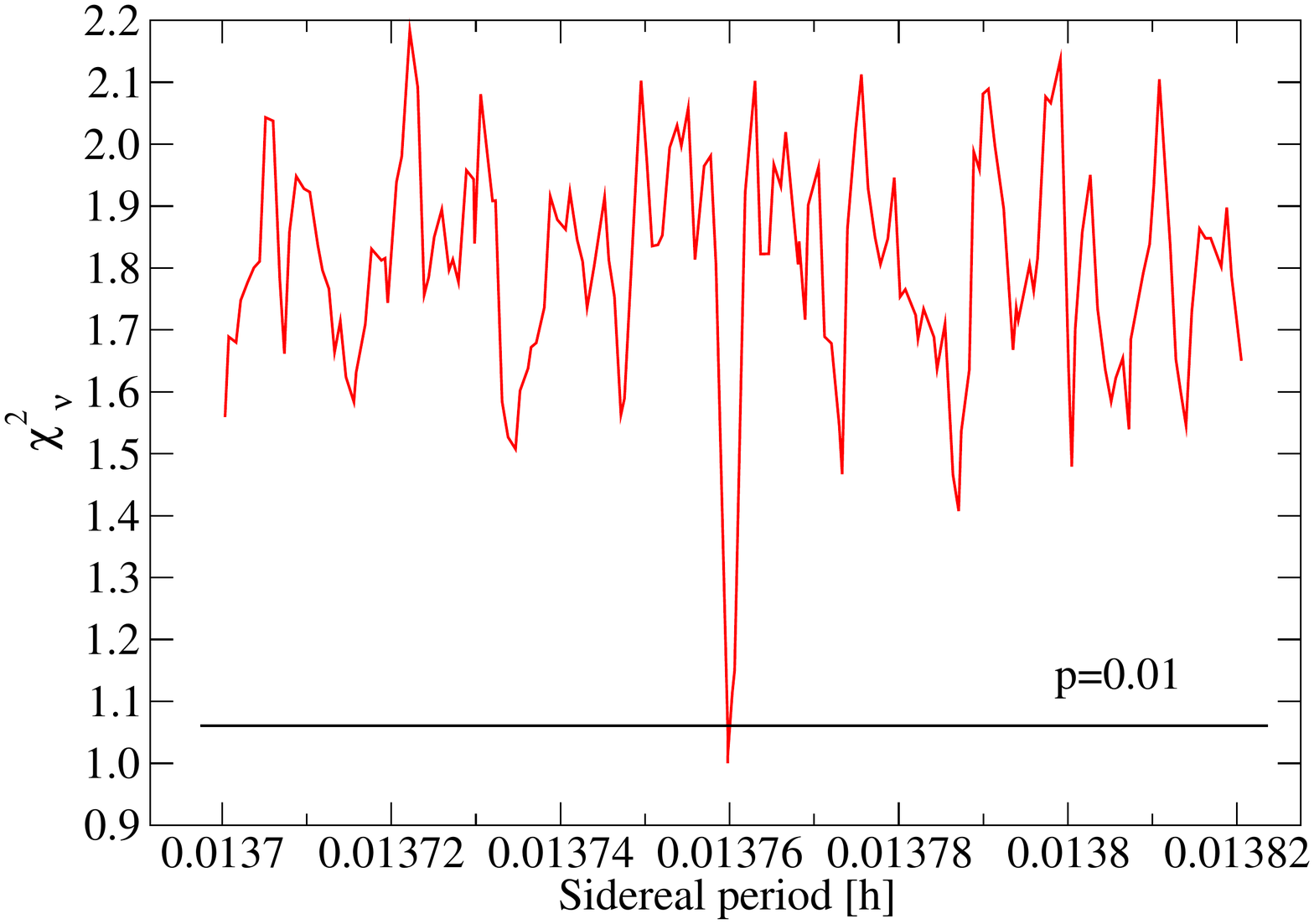}}
\caption{Results of the search for the sidereal period of rotation for the
$P_{2}$ solution.  The search interval has been defined by the range of
change of the synodic period. On the vertical axis there is a value of 
$\chi^{2}_{\nu}$ ($\chi^2$ per degree of freedom). A horizontal line indicates 
a p-value of 0.01. The only statistically significant minimum (which reaches
below p=0.01) is $P_{\mathrm{sid}}=0.013760\pm 0.000001\,\mathrm{h}$.}
\label{P2_sid}
\end{figure}

Fig.~\ref{P2_sid} presents a periodogram for the sidereal period, with the
best result obtained for $P_{\mathrm{sid}}=0.013760 \pm
0.000001\,\mathrm{h}$.  There were two solutions for the ecliptic
coordinates of the spin axis, A: $\lambda_1=57\degr \pm 10\degr,
\beta_1=29\degr \pm 10\degr,$ and B: $\lambda_2=67\degr \pm 10\degr,
\beta_2=-40\degr \pm 10\degr$, both with the same sidereal period. 
The obliquities resulting from those pole orientations are: 
$\epsilon_1=54\degr \pm 10\degr$ and $\epsilon_2=123\degr \pm 10\degr$.
They are far away from the asymptotic states ($\epsilon=0\degr, 180\degr$)
predicted by the theory.

\begin{figure}[!t]
\resizebox{\hsize}{!}{\includegraphics[clip]{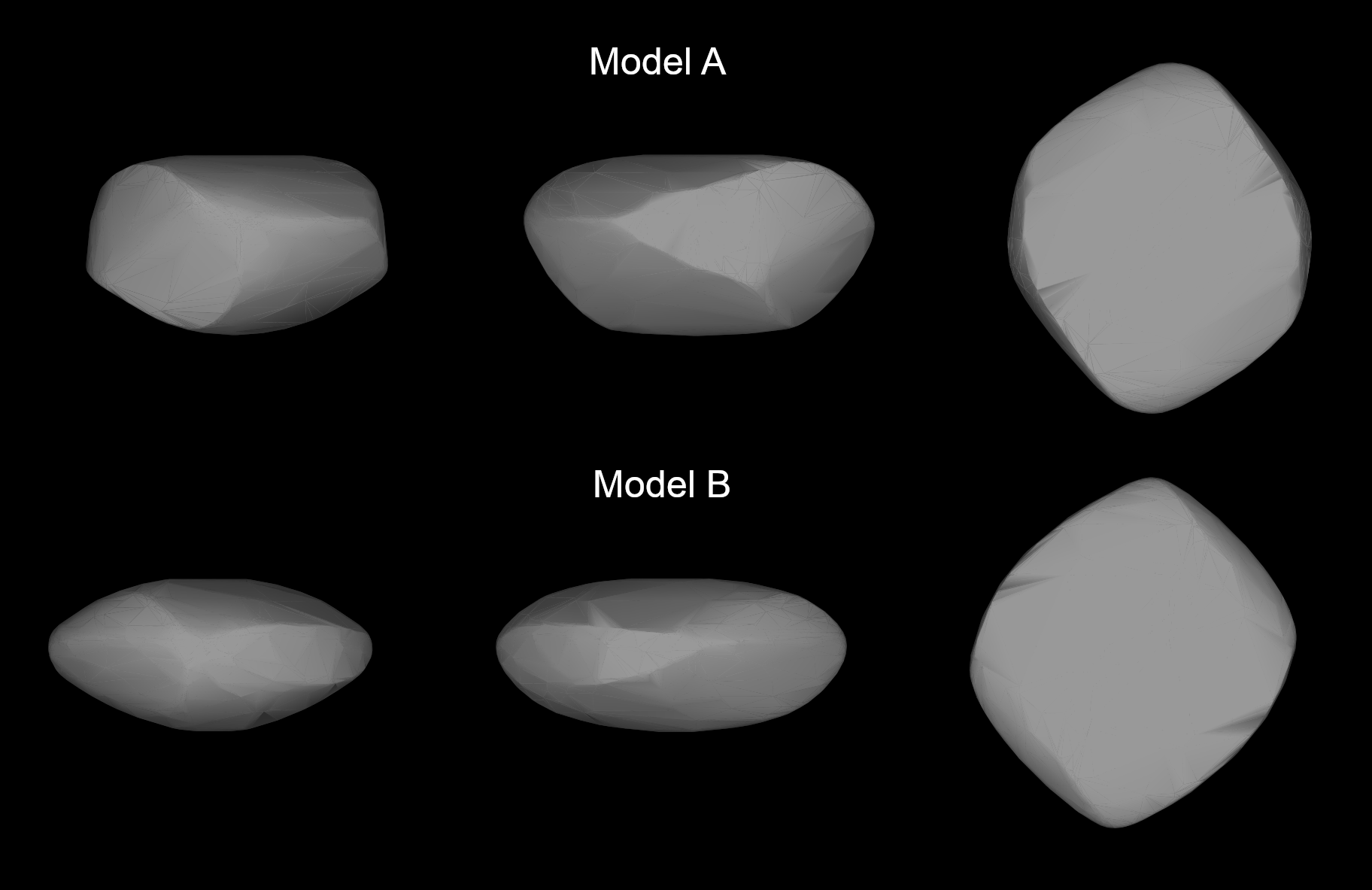}}
\caption{Projections of convex shapes obtained for two pole solutions.
The upper row is for pole A, the lower one for pole B. In the left column we
can see a projection onto the x-z plane, in the middle column: onto the y-z
plane, while at the right there are projections on the x-y plane (top
views, along the rotation axis).} 
\label{convex_shapes}
\end{figure}

\begin{figure*}[!tbh]
\centering
\resizebox{0.9\hsize}{!}{\includegraphics[clip]{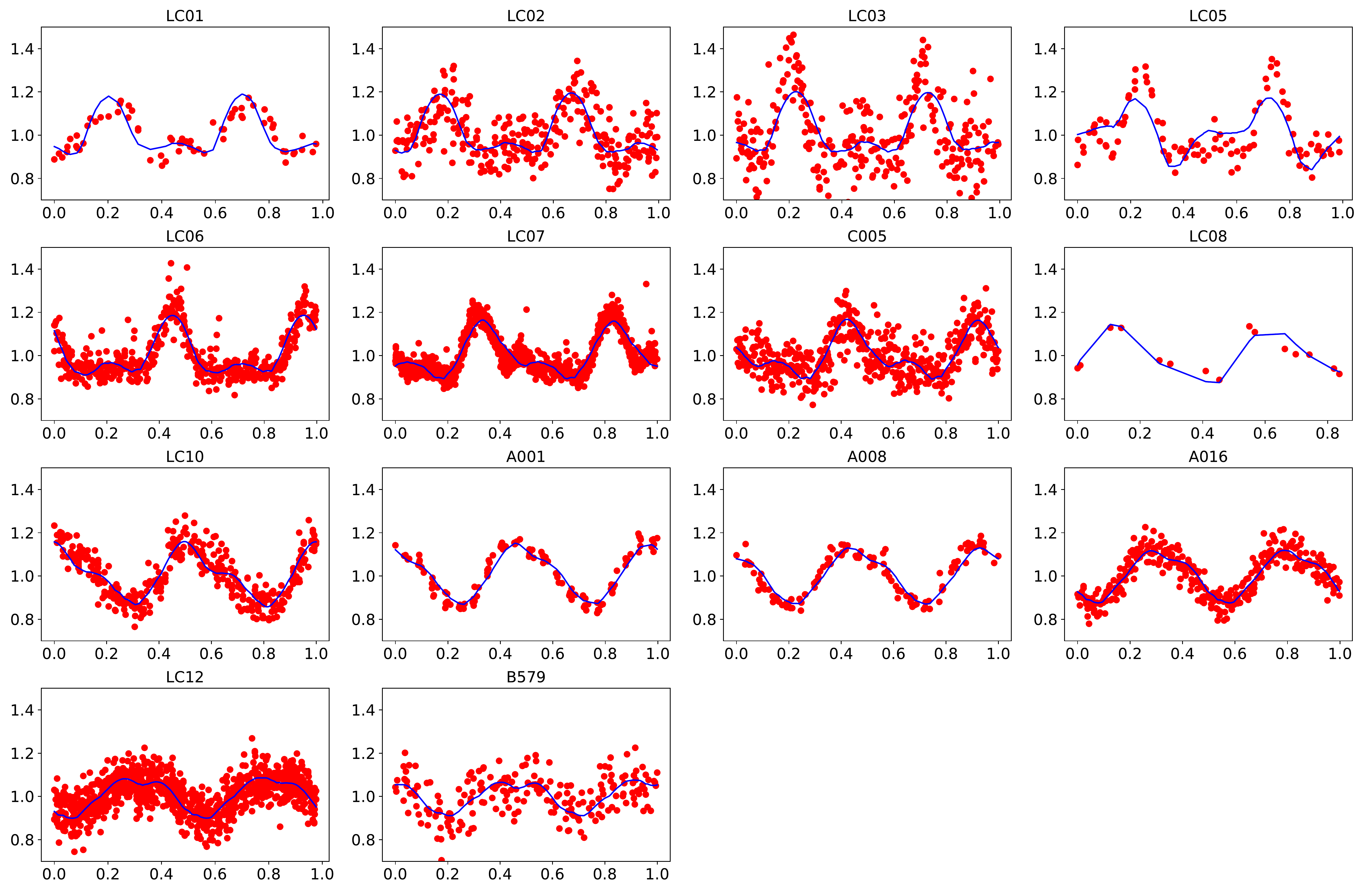}}
\caption{A superposition of the model~A lightcurves (continuous line) on the
data points obtained from observations.  All plots are drawn in the same
coordinates: the x-axis is the rotation phase (in the range from 0 to 1),
the y-axis refers to the flux scaled to unity for the average brightness. 
Note that our model convex have problems with recreating sharp brightness
maxima in LC03, LC05, and LC06 lightcurves.  They are most likely caused by
concavities on the asteroid surface.}
\label{data_vs_model}
\end{figure*}

As it often happens, the positions are symmetric with respect to the
ecliptic plane (within the specified uncertainties).  Convex shapes
resulting from these pole positions are presented in
Fig.~\ref{convex_shapes}, and a fit of the modelled lightcurves to the data,
for pole A, in Fig.~\ref{data_vs_model}.  The fits obtained for solution B
are very similar.  For both models the percentage of dark facets was smaller
than 0.5.  Note that in the final solution we did not use LC09, which is
very close in time to LC10, and did not give any new information to the
optimization algorithm.

Scanning the whole celestial sphere for other solutions we encountered
cases, where the fit of the model lightcurves to the observed ones was
satisfactory, but the convex shapes were not physical.  For example, the
body was significantly elongated in the c-axis direction (which was assumed
to be the axis of rotation), the shape was unrealistically flat (pan-cake
like), or it had more than 5\% of dark facets on the surface.  The latter
would imply surface albedo variegation, which is difficult to accept for such
small objects. 

The same happened when we started with the
$P_1$ synodical period. Folding all data with $P_1$ we obtained a set of
mono-modal composite lightcurves, well fit by the Fourier series. However,
when we used such data for lightcurve inversion, we obtained unrealistic
solutions. Based on this, we reject $P_1$ as a possible period of
rotation (with the assumption of convex shape and the lack of albedo
variegation on the surface).

\section{Magnitude-phase curve}

There are several problems which make it difficult to derive the
magnitude-phase angle function (hereafter: phase function) for NEAs. 
Firstly, they are seldom observed close to opposition, so the non-linear
part of the phase function cannot be properly determined.  Secondly, during
their passage close to Earth, the aspect angle (the angle formed by the
object's spin axis and the direction towards the observer) can change
significantly, influencing the magnitude-phase angle relationship.  Thirdly,
during astrometric observations of NEAs (which are used routinely by the
Minor Planet Center\footnote{https://www.minorplanetcenter.net/} (hereafter
MPC) for determination of the phase function), asteroid magnitudes from
different phases of rotation are reported and the phase curve is distorted
even more by the rotational brightness variations.

In the case of 2021~DW$_1$ we observed its full lightcurves and were able to
determine its rotationally averaged magnitudes.  This removed the third
obstacle mentioned above.  Since our observations were done in different
filters (and often were ''unfiltered''), for the purpose of photometric
calibration we used the PanSTARRS catalogue \citep{Tonry+2012} standards of
solar colours.  We selected the best lightcurves reported in
Table~\ref{AspectData}, and calibrated them in the SDSS~r band.  For each
lightcurve we used the Fourier fit to compute its mean magnitude.  We also
used the data from 7~March (lightcurve LC013), which were quite noisy for
the convex shape modelling, but gave a good rotationally averaged magnitude.

In the process we had to discard some results, where calibration was
inaccurate.  We also noticed that our data were strongly affected by aspect
changes during the time, when the asteroid was changing its ecliptic
latitude from $\beta=-25\degr$ to $\beta=+20\degr$.  For this reason we used
only lightcurves observed from 4~Mar, 2~UTC to 7~Mar, 8:45~UTC.  After
scaling the magnitudes to unit distances from Earth and Sun, we tried to fit
them with a standard H, G phase function.  Our standard program for doing
that, with an advanced nonlinear minimization, failed, so we used a simpler
approach changing the H and G values in some intervals, fitting the H,~G
function to the data, and selecting the result with the smallest residuals. 
Unfortunately, the best fit was obtained for $G=1.1$ which has no
physical meaning because the typical values of G for asteroids fall between 
0.0 and 0.5. The absolute magnitudes $H_{\mathrm{r}}$
obtained for $G=0.0$ and $G=0.5$, where $H_{\mathrm{r}}=24.00$ and  
$H_{\mathrm{r}}=24.98$, respectively. To translate them to $H$ values in 
the V band (which is the standard way of
reporting absolute magnitudes), one can increase them by 0.21~mag.  
This value can be obtained using the transformation from $r$ to $V$, provided by
\citet{Tonry+2012}.

\begin{figure}[!t]
\resizebox{\hsize}{!}{\includegraphics[clip]{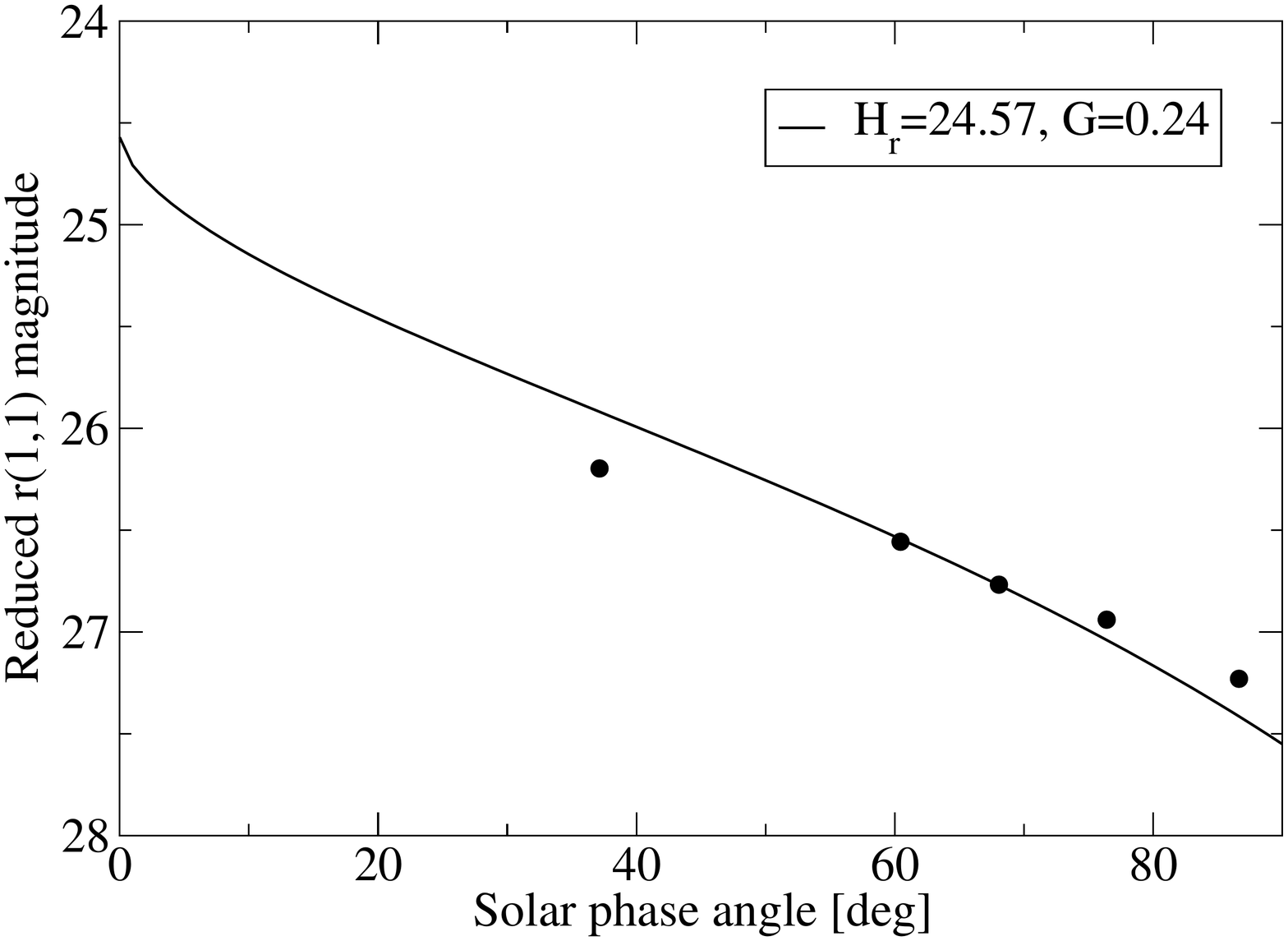}}
\caption{Phase curve of 2021~DW$_1$ and the H,~G function fit. 
During fitting, the $G$ parameter was kept constant at 0.24 (a value typical
for S-type asteroids).  Note that the magnitudes used are in the SDSS~r
band.  The obtained $H_\mathrm{r}=24.57$, after conversion to the Johnson~V
band, gives $H_{\mathrm{V}}=24.8$}
\label{PhaseCurve}
\end{figure}

The absolute magnitude for 2021~DW$_1$, derived by the MPC from 284 less
accurate, astrometric observations, is $H_{\mathrm{MPC}}=25.02$.  It is obtained assuming $G=0.15$.  To get a final value for the absolute
magnitude of 2021~DW$_1$ we used the fact that (as we describe in the next
section) it is an S-type asteroid, for which $G=0.24\pm 0.11$
\citep{Warner+2009}.  With this assumption we got 
$H_{\mathrm{r}}=24.57$ which, after conversion to the V band, is 
$H_{\mathrm{V}}=24.8$ (Fig.~\ref{PhaseCurve}).

Since in our computation of $H$ systematic effects dominate over the
statistical ones, we are not able to derive a standard ''one $\sigma$'' error for
it. Instead, we estimate the accuracy of the obtained absolute magnitude
by a maximum uncertainty of $\Delta H=0.5$~mag. 

\section{Colour indices and taxonomy}

On 5 March we observed the asteroid in three Sloan g, i, and z filters (see the rows in Table~\ref{AspectData} with the codes CI01-3). For each of them, a serie of 4~s exposures where obtained with the telescope tracking on the asteroid. 

In the case of the observations in i and g filters, we had no problems with data reduction, while the exposures in the infrared z-filter left interference fringes on the CCD frames. The reduction of fringing is possible by subtracting the so-called fringe map, which is obtained by averaging the images of the blank fields (fields with a few visible stars). We could not make blank field observations until 30 April. Unfortunately, it turned out that the fringe pattern changed too much from 5 March till 30 April to use the obtained fringe map for the reduction.

The non-sidereal telescope tracking ensured an almost constant position of the asteroid image in relation to the fringe structure, thanks to which it was possible to avoid the influence of fringing on instrumental measurements of the asteroid. The problem was with the comparison stars, which moved quickly in relation to the fringe structure. To minimize this effect, we selected just one PanSTARRS comparison star, for which the fringing had little effect on its brightness (standard deviation of brightness measurements $\sigma = 0.02 \,\mathrm{mag}$). We used this star to calibrate the z filter magnitude of the asteroid. 

Next, using PerFit, we phased the g, i, and z lightcurves with the reference VR lightcurve, obtaining one composite lightcurve. For this purpose, we used part of the observations made in the VR filter on the same telescope just before the exposures in the g filter, and for control, 40 minutes after the exposures in the z filter. In both cases the aspect data were practically the same. The obtained values of $VR-g$, $VR-i$ and $VR-z$ magnitude shifts gave the following colour indices:
$g-i=0.79\pm 0.01\,\mathrm{mag}$,
$i-z=0.01\pm 0.02\,\mathrm{mag}$,
$z-g=0.01\pm 0.02\,\mathrm{mag}$.

\begin{figure}[!t]
\resizebox{\hsize}{!}{\includegraphics[clip]{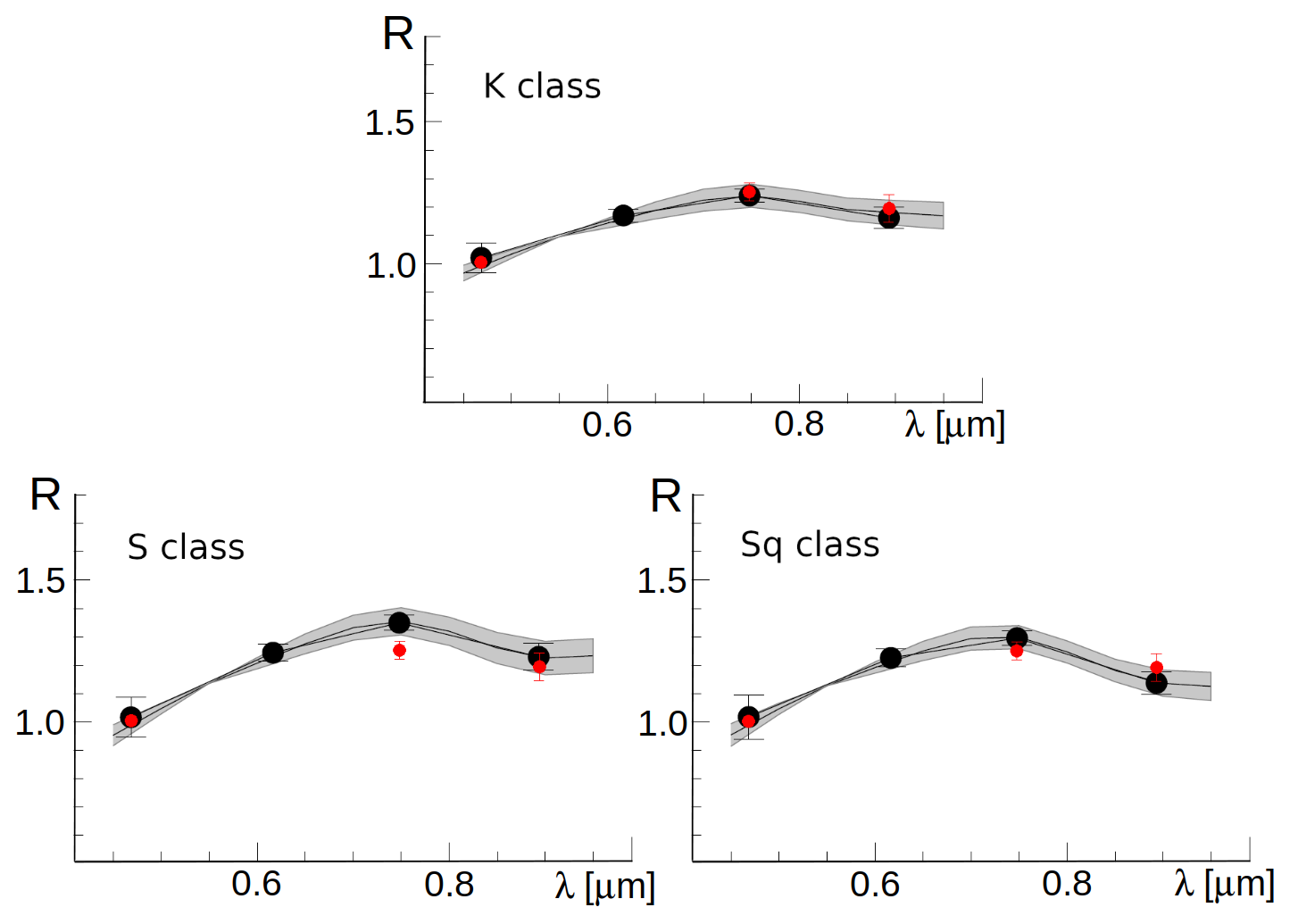}}
\caption{Plots of reflectivities $R_{\mathrm{g}}$, $R_{\mathrm{r}}$,
$R_{\mathrm{i}}$, and $R_{\mathrm{z}}$ of three asteroid taxonomic classes (black
circles), which are most similar to the reflectivities of 2021~DW$_1$ (red
circles). All reflectivities are normalized to the g band, and the
grey areas denote a dispersion of spectra in a given asteroid class.
Data for the taxonomic classes were taken from \cite{DeMeo+2013}.}
\label{Rvslambda}
\end{figure}

To determine the taxonomic type of 2021~DW$_1$, we converted the colour indices to the reflection coefficients $R_{i}$ and $R_{z}$, normalized to the reflection values in the g~band, using the formula given in \cite{DeMeo+2013}:
\begin{equation}
R_{f}=10^{-0.4[(f-g)-(f_{\odot}-g_{\odot})]}
\label{DeMeo}
\end{equation}
where $g$ and $g_{\odot}$ are the g magnitudes of the asteroid
and Sun, and $f$ and $f_{\odot}$ are the magnitudes in some other band.  In
Eq.~\ref{DeMeo} we assumed the solar colours $(i-g)_{\odot} = -0.55\pm 0.03
\,\mathrm{mag}$ and $(z-g)_{\odot} = -0.61 \pm 0.04 \,\mathrm{mag}$
\citep{Holmberg+2006}, used our colour indices, and obtained the following
reflectivities: $R_\mathrm{g}\equiv 1$, $R_{\mathrm{i}}=1.25\pm 0.03$, and
$R_{\mathrm{z}}=1.19 \pm 0.05$.  Next, we compared them with spectra of
different taxonomic classes, as given by \citet{DeMeo+2013}.  A good match
was found for the S, Sq and K spectra (Fig.~\ref{Rvslambda}), with the Sq
class beeing the best.



Since K-class asteroids are typical mainly for the Eos family, and are rare
in the asteroid population, the confirmation of this possibility would
require a good quality spectrum of 2021~DW$_1$.  In our campaign we planned
to get such a spectrum, but a fast movement of the object on the sky made it
impossible even for the robotic spectrograph mounted on the 2.0-m~LCO
telescope.  Since K-class objects are very rare among asteroids, our current
conclusion is that 2021~DW$_1$ is a typical S-class NEA.

\section{Physical characteristic of 2021 DW1}

Having obtained the absolute magnitude $H_\mathrm{V}=24.8\pm 0.5$, and an estimate of
the geometric albedo $p_{\mathrm{V}}=0.23\pm 0.02$, we are now able to derive the
asteroid effective diameter $D_{\mathrm{eff}}$\footnote{Diameter of a sphere
which has the same brightness as the rotationally averaged magnitude of the
asteroid -- assuming a zero solar phase angle.}.  For that, we used a
standard equation $D_{\mathrm{eff}}=1329\times
10^{(-H_{\mathrm{V}}/5)}\times p_{V}^{(-1/2)}$ from Fowler and Chillemi
(1992), and obtained $D_{\mathrm{eff}}=30\, \mathrm{m}$. To get its uncertainty we
first converted the statistical error $\sigma_{\mathrm{pV}}$ into a maximum
one: $\Delta p_{\mathrm{V}} = 3\times \sigma_{\mathrm{pV}}$. Then we used both  
$\Delta H=0.5$ and $\Delta p_{\mathrm{V}}=0.06$ to compute the
maximum uncertainty $\Delta D=10\,\mathrm{m}$. With that we finally get: 
$D_{\mathrm{eff}}=30\pm 10\,\mathrm{m}$.

\begin{figure}[!t]
\resizebox{\hsize}{!}{\includegraphics[clip]{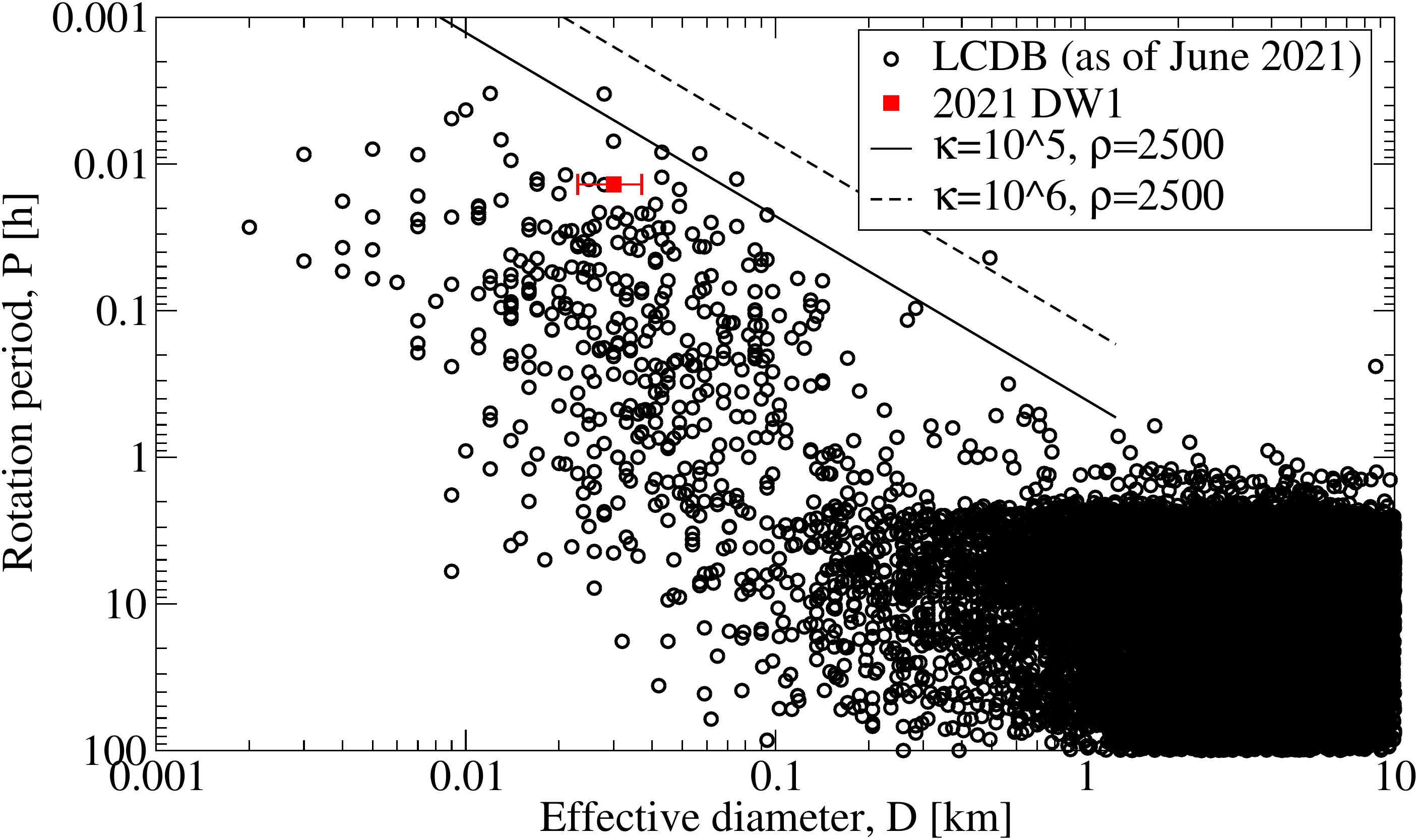}}
\caption{Position of 2021~DW$_1$ on the $\log D_{\mathrm{eff}} - \log P$ plot. 
The open circles were taken from the LCDB, while sloped lines indicate
minimum allowable periods (at a given diameter), computed with two different
tensile strength coefficients ($\kappa=10^5$ and $\kappa=10^6 \,\mathrm{N}\,
\mathrm{m}^{3/2}$), and a density of $\rho=2500\,\mathrm{kg m^{-3}}$.  A red
square indicates the position of 2021~DW$_1$ with the error bars being a
measure of the maximum uncertainty.}
\label{logD_logP}
\end{figure}

One of the methods for studying physical properties of asteroids is to plot
their rotation periods versus effective diameters.  On Fig.~\ref{logD_logP}
we present results, taken from the last issue of the Light Curve Data Base
(LCDB)\footnote{https://minplanobs.org/mpinfo/index.php}
\citep{Warner+2009}.  The plot includes not only VSAs, but also larger
objects -- both NEAs and Main Belt Asteroids -- up to 10 kilometres in
diameter.  What is clearly visible in this figure is that most asteroids
with $D_{\mathrm{eff}} > 1\,\mathrm{km}$ (in the lower-right corner of the
plot) have periods $P>2.2\,\mathrm{h}$, while most of VSAs display much
faster rotation.  This is possible because VSAs are held together by tensile
strength rather than gravity.  However, there is a lower limit to their
periods, set by centrifugal forces.  \citet{Holsapple2007} derived an
approximate formula, which makes it possible to draw, on the $\log
D_{\mathrm{eff}}$ -- $\log P$ plot, a line of minimum allowable
periods (which we will call critical periods, $P_{\mathrm{c}}$) for the
asteroid in the strength regime.  We used a slightly modified equation for
$P_{\mathrm{c}}$, given by \citet{Kwiatkowski+2010a}, to compute such lines
on Fig.~\ref{logD_logP} for two tensile strength coefficients ($\kappa=10^5$
and $\kappa=10^6 \,\mathrm{N}\, \mathrm{m}^{3/2}$).  A justification of our
choice of $\kappa$ can be found in \citet{Kwiatkowski+2010a}.  We also
assume a typical angle of friction $\phi=40\degr$ \citep{Richardson+2005}, a
density $\rho=2500\, \mathrm{kg m^{-3}}$, and the triaxial ellipsoid
approximation of the 2021~DW$_1$ shape given by $c/a=0.48, b/a=0.88$.  The
last two parameters are derived by measuring the extension of the asteroid
shape along the x,~y,~z axes, and computing the average from Model~A and B. 
Note that the position of the $P_{\mathrm{c}}$ line is most
sensitive to $\kappa$ and $\rho$.

On Fig.~\ref{logD_logP} we also mark a position of 2021~DW$_1$ with its
uncertainty.  Since we ruled out the $P_1$ solution for the rotation period,
all uncertainty now lies along the diameter axis. As can be seen, 2021~DW1 
is far from the lines denoting the rotational fission. In the future,
depending on the YORP-TYORP cycles, it will move vertically up (or down), or
may be trapped in some equilibrium point, where its period and spin axis 
will not change.

\section{Conclusions}

We have derived the spin axis coordinates for
2021~DW$_1$.  In the ecliptic reference frame, the two solutions are: (A)
$\lambda_1=57\degr \pm 10\degr, \beta_1=29\degr \pm 10\degr$, and (B)
$\lambda_2=67\degr \pm 10\degr, \beta_2=-40\degr \pm 10\degr$, with obliquities $\epsilon_1=54\degr\pm10\degr$ and $\epsilon_2=123\degr \pm 10\degr$,
respectively. It shows that the spin axis of 2021~DW1 is far from
the asymptotic states of $\epsilon=0\degr, 180\degr$ predicted by
simulations and theory \citep{Capek+2004, Golubov+2021} for asteroids with
high thermal conductivity.  Interestingly, both $\epsilon_1$ and
$\epsilon_2$ have the same values as the obliquities, at which the period
change component of YORP vanishes, as shown by Eq.~13 in
\citet{Golubov+2021}.  The same effect has been observed in numerical
simulations by \citet{Capek+2004} (their Figs.~6-8). However, this is only a
temporary situation because the obliquity change component of YORP at
those two positions is still significant. After some time, as the obliquity 
continues to change, the rotation period can be altered again.

We also obtained other physical parameters of 2021~DW$_1$: a sidereal period
$P_{\mathrm{sid}}=0.01376\pm 0.00001\,\mathrm{h}$, the parameters of its 
magnitude-phase
function $H=24.8\pm 0.5\,\mathrm{mag}$ and $G=0.24$.  The asteroid colour
indices are $g-i=0.79\pm 0.01\,\mathrm{mag}$, and $i-z=0.01\pm
0.02\,\mathrm{mag}$ which indicates an S taxonomic class, with an average
geometric albedo $p_V=0.23\pm 0.02$.  The asteroid effective diameter,
derived from $H$ and $p_V$, is $D_\mathrm{eff}=30\pm 10\,\mathrm{m}$.

Unfortunately, no new observations of this object will be possible in the
near future.  According to the JPL~Horizons service, the only close approach
of 2021~DW$_1$ to Earth in the next 100 years will take place on 20 February
2047.  At this time, however, the asteroid will reach the maximum brightness
of $V=21\,\mathrm{mag}$ and, with such rapid rotation, would require the
biggest telescopes for observations.

\section{Acknowledgements}

T.~Kwiatkowski, A.~Kryszczyńska, and D.~Oszkiewicz were supported by a grant
No.2017/25/B/ST9/00740 from the National Science Centre, Poland.  T.~Kim was
supported by the Korea Astronomy and Space Science Institute under the R\&D
program (Project No.  2020-1-600-05) supervised by the Ministry of Science
and ICT (MSIT), and the National Research Foundation of Korea (NRF) Grant
No.  2020R1A2C3011091, funded by MSIT. The work by T.~Santana-Ros was carried out through a grant APOSTD/2019/046 by
Generalitat Valenciana (Spain). He was also supported by the MINECO (Spanish Ministry of Economy) through a grant RTI2018-095076-B-C21
(MINECO/FEDER, UE). During observations the NASA
JPL~Horizons service was used extensively. We would like to thank the 
reviewer, David Vokrouhlick{\'y}, for his useful comments which improved the
article.

\bibliographystyle{aa}
\bibliography{dw1}

\end{document}